\documentclass{article}
\usepackage{graphicx}  % standard LaTeX graphics tool;
\usepackage{amsmath}   % For getting proper math eqns
\usepackage[noadjust]{cite}
\usepackage{amssymb}   % Used for getting various symbols eg. gtrsim, lesssim etc.
\usepackage{bm} % For bold math (esp of lower greek letters)
\usepackage{dcolumn}% Align table columns on decimal point
\usepackage{color}
\usepackage{mathrsfs}
\usepackage{amsfonts}
\usepackage{varioref}
\usepackage{braket}
\usepackage{subcaption}
\usepackage{verbatim}
\usepackage{rotating} 

\RequirePackage[colorlinks,citecolor=blue,urlcolor=magenta,linkcolor=blue]{hyperref}
\def\eq#1{{Eq.~(\ref{#1})}}

\def\eq#1{{Eq.~(\ref{#1})}}

\def\frab#1#2{\left(\frac{#1}{#2}\right)}

\def\ket#1{|#1\rangle}                    %%%%   ket 
\def\bk#1#2#3{{\langle #1|#2|#3\rangle}}  %%%%   bracket
\def\tv#1{\bm{#1}_\perp}

\title{Exploring the Rindler vacuum and the Euclidean Plane}
\author{Karthik Rajeev\footnote{karthik@iucaa.in}~and~
	T. Padmanabhan\footnote{paddy@iucaa.in}\\
	{\small{IUCAA, Post Bag 4, Ganeshkhind, Pune University Campus, Pune 411007, India.}}}
\date{26 May, 2020}  
\begin{document}
	\maketitle
	\begin{abstract}
		In flat spacetime, two inequivalent vacuum states which arise rather naturally are the Rindler vacuum $\ket{\mathcal{R}}$ and the Minkowski vacuum $\ket{\mathcal{M}}$. We discuss several aspects of the Rindler vacuum, concentrating on the propagator and Schwinger (heat) kernel defined using $\ket{\mathcal{R}}$, both in the Lorentzian and Euclidean sectors. We start by exploring an intriguing result due to Candelas and Raine [P. Candelas and D. J. Raine, J. Math. Phys., 17,
		 2101(1976)], viz., that $G_{\mathcal{R}}$, the Feynman propagator corresponding to $\ket{\mathcal{R}}$, can be expressed as a curious integral transform of $G_{\mathcal{M}}$, the Feynman propagator in $\ket{\mathcal{M}}$. We show that this relation follows from the well-known result that, $G_{\mathcal{M}}$ can be written as a periodic sum of $G_{\mathcal{R}}$, in the   Rindler time $\tau$,  with the period (in proper units)  $2\pi i$. We further show that, the integral transform result holds for a wide class of pairs of bi-scalars $\left\{F_{\mathcal{M}},F_{\mathcal{R}}\right\}$, provided $F_{\mathcal{M}}$ can be represented as a periodic sum of $F_{\mathcal{R}}$ with period $2\pi i$. We provide an explicit procedure to retrieve $F_{\mathcal{R}}$ from its periodic sum $F_{\mathcal{M}}$, for a wide class of functions.  An example of particular interest is the pair of Schwinger kernels $\left\{K_{\mathcal{M}},K_{\mathcal{R}}\right\}$,  corresponding to the Minkowski and the Rindler vacua. We obtain an explicit expression for 
		$K_{\mathcal{R}}$ and clarify several conceptual and technical issues related to these biscalars both in the Euclidean and Lorentzian sector. In particular, we address the issue of retrieving the information contained in all the four wedges of the Rindler frame in the Lorentzian sector, starting from the Euclidean Rindler (polar) coordinates. This is possible but requires four different types of analytic continuations, based on one unifying principle. Our procedure allows generalisation of these results to any (bifurcate Killing) horizon in curved spacetime.

	\end{abstract}

	\section{Introduction and Motivations}
	
In standard quantum field theory, the Fock basis is introduced by identifying the creation and annihilation operators, followed by defining the vacuum $\ket{0}$ as the unique state annihilated by the latter, and building multi-particle states via repeated action of the creation operators on $\ket{0}$. For instance, for a real Klein-Gordon (KG) scalar field $\Phi$, the creation operators ($a^{\dagger}_{j}$) and annihilation operators ($a_{j}$) are identified as the operator-valued coefficients in the following expansion of the corresponding Heisenberg operator:
	\begin{align}
	\Phi=\sum_{j}\left[a_{j} \phi_{j}+a_{j}^{\dagger}\phi_{j}^*\right]
	\end{align}
	where, $\phi_{j}$ are the positive frequency modes such that $\{\phi_{j},\phi^*_{j}\}$ is a complete orthonormal (under the KG inner product) basis set for the solutions to the KG equation. The solutions $\phi_{j}$, in turn, are defined as those that approximate a positive energy mode near an appropriate time $t_0$, which is usually taken to be the asymptotic past or future.
	As is well known, such procedures are not unique, and one can easily construct an inequivalent class of creation/annihilation operators, thereby leading to different vacuum states \cite{Fulling:1972md}. 
	
	One such situation --- extensively studied in literature --- corresponds to the notions of Rindler vacuum and Minkowski vacuum \cite{Unruh:1976db,Birrell:1982ix,Wald:1995yp,Lee:1985rp,Unruh:1983ac}, which arise along the following lines.
	In a $D$-dimensional flat spacetime, introduce the standard Lorentzian coordinates $x^a = (t, x, \bm{x}_\perp)$. The $x-t$ plane is divided into four wedges $(R,L,F,P)$ by the $x=\pm t$ planes in the standard manner (see Figure 1). Introduce the Rindler coordinates $(\tau, \rho, \bm{x}_\perp)$ in the right ($R$) and the left ($L$) wedges   with, for e.g., $x=(\rho/g) \cosh (g\tau), t = (\rho/g) \sinh (g\tau)$ in $R$ and $x=-(\rho/g) \cosh (g\tau), t = -(\rho/g) \sinh (g\tau)$ in $L$, with similar transformations in other wedges \cite{Rindler:1966zz}. (We assume that $\rho>0$ and $-\infty<\tau<\infty$.)
	The metric is static with respect to both $t$ and $\tau$ coordinates and hence one can find mode functions which are positive frequency with respect to $t$ or with respect to $\tau$. The corresponding creation/annihilation operators can be used to define the Minkowski vacuum $\ket{\mathcal{M}}$ and the Rindler vacuum $\ket{\mathcal{R}}$. One can then build standard QFT based on these two vacua and study their inter-relationship. In particular, one can define the Minkowski and Rindler propagators by the standard procedure: 
	\begin{equation}
	G_\mathcal{M} (x_2,x_1) \equiv  \bk{\mathcal{M}}{T_t[\phi(x_2)\phi(x_1)]}{\mathcal{M}}; \qquad   G_\mathcal{R} (x_2,x_1) \equiv \bk{\mathcal{R}}{T_\tau [\phi(x_2)\phi(x_1)]}{\mathcal{R}}
	\end{equation} 
	where the subscripts on $T$ indicate the time-coordinate used for time-ordering. 
	
	\begin{figure}[h!]
		\centering
		\includegraphics[scale=.27]{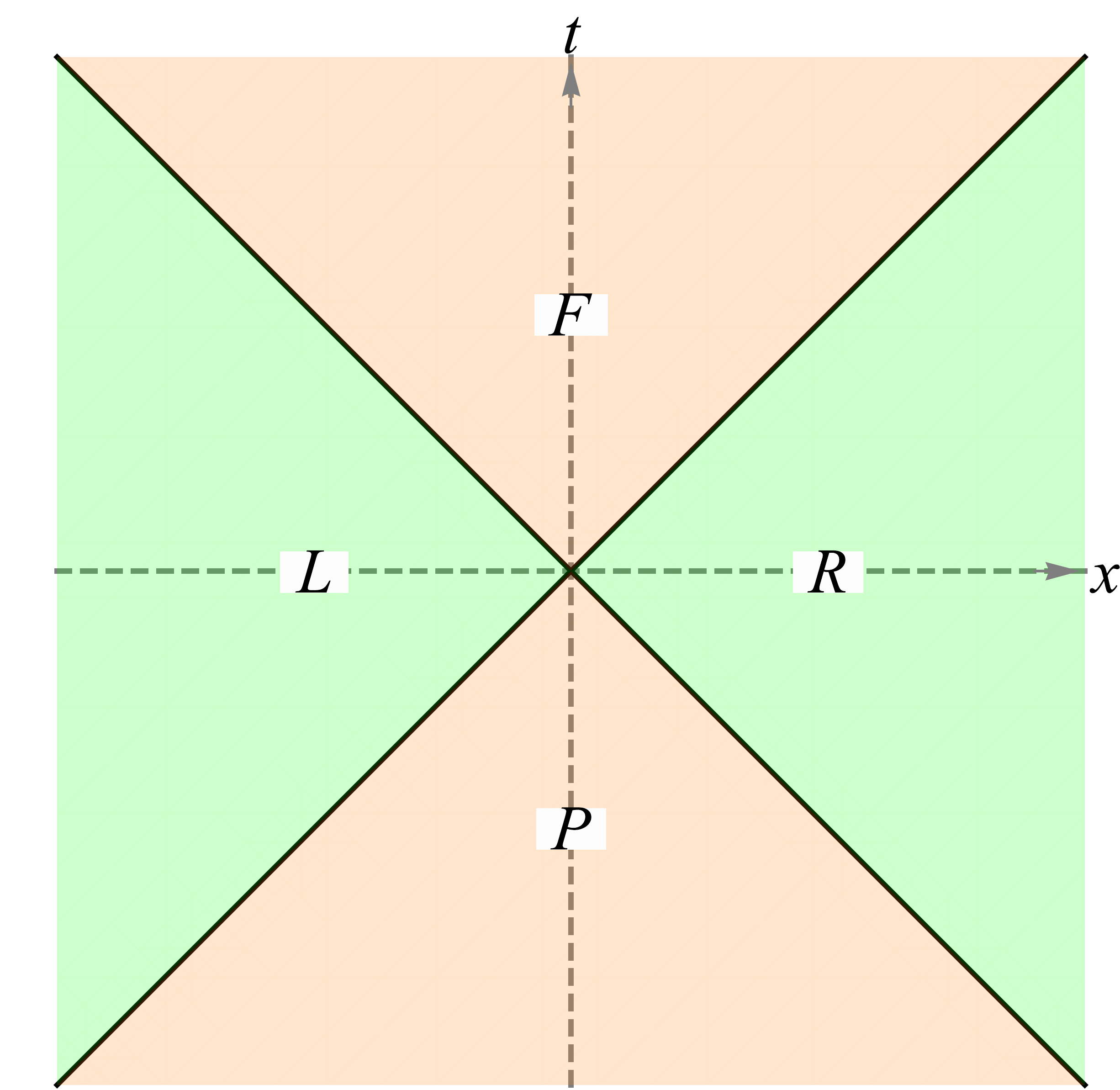}
		\caption{The four `wedges' of the $x-t$ plane of Minkowski spacetime}
		\label{fig:wedges}
	\end{figure} 
	
	It is also well known that $\ket{\mathcal{M}}$ behaves like a thermal state filled with Rindler particles at  the temperature $T= (g/2\pi)$, where $g$ is the proper-acceleration of the Rindler observer moving along the trajectory $\rho=1$. This thermality  implies that 
	the Minkowski propagator $G_\mathcal{M}$ can be thought of as a `thermalised' version of the Rindler propagator $G_{\mathcal{R}}$ in the following sense \cite{Troost:1978yk}: 
	\begin{equation}
	G_{\mathcal{M}} (i\tau) = \sum_{n=-\infty}^\infty G_\mathcal{R} \left(i\tau + i\,2\pi n\,g^{-1}\right)
	\label{thermaltp}
	\end{equation} 
	This equation is to be understood as  summarizing  a three-step process: (i) Take the function $G_{\mathcal{R}}(\tau) $ and analytically continue it to the Euclidean time $\tau_E \equiv i\tau$. (ii) Construct the sum in the right hand side by changing $\tau_E \to \tau_E +2\pi ng^{-1}$ and summing over all $n$. (iii) Analytically continue back to $\tau$.  The result is formally expressed by  \eq{thermaltp}. Hereafter, we will say that a function $f_{\mathcal{M}}(\tau)$ is a \textit{thermalised version} of a function $f_{\mathcal{R}}(\tau)$ when the two functions are related by this procedure.  For convenience, we shall henceforth work in a system of units in which $g=1$.
	
	\subsection{Motivation 1: Probing an intriguing relation}  \label{mot1}

	There is, however, another intriguing relationship between $G_{\mathcal{M}}$ and $G_{\mathcal{R}}$, which has received very little attention in the literature.  It turns out that, for events $(x_1, x_2)$ in $R$, there is a curious relation between $G_{\mathcal{R}}$ and $G_{\mathcal{M}}$ given by:
	\begin{equation}
	G_{\mathcal{R}}(x_1,x_2) = G_{\mathcal{M}}[\sigma(x_1,x_2)] - \int_{-\infty}^\infty d\lambda\ \frac{G_{\mathcal{M}}[\sigma(x_1,x_2^{(r)}(\tau_2-\lambda))]}{\pi^2+(\lambda-\tau_1)^2}
	\label{one}
	\end{equation} 
	where $\sigma^2(x,y)$ is the square of the invariant distance between the two events and, the event $x_2^{(r)}(\tau)$ is defined through the relation $x_2^{(r)}(\tau)=x_2(\tau\pm i\pi)$. Geometrically, one can interpret $x_2^{(r)}(\tau)$ as the `reflection' of $x_2(\tau)$ about the origin of the $x-t$ plane, as shown in Figure 2. This also implies,
	\begin{equation}
	\sigma^2(x_1,x_2^{(r)}(\tau_2-\lambda)) = \rho_1^2 +\rho_2^2 + 2 \rho_1\rho_2 \cosh(\lambda-\tau_2) + (\Delta \tv{x} )^2 = \sigma^2 (\tau_1,\tau_2 \pm i\pi)
	\end{equation} 
	In the last expression, we have only displayed the dependence on the Rindler time coordinate.
	The relation \eq{one} was first obtained in 1976 by Candelas and Raine  \cite{cr} by a fairly lengthy, detailed, computation.  In the last four decades this result has received very little attention or elaboration in the literature  --- e.g., the reviews \cite{takagi,Crispino:2007eb} do not even mention it ---  though the structure of $G_{\mathcal{R}}$ has been investigated by several authors (for example, \cite{linet,Moretti:1995fa,Troost:1978yk,FULLING1987135} etc.) in the intervening years. In fact, the only two other papers in which \textit{we could find} this result quoted briefly, without any elaboration, were Refs. \cite{jarmo,long-shore}. 
	This relation, however, is rather intriguing because of the following features:
	
	(a) The second term in the right hand side of \eq{one} uses $G_{\mathcal{M}}$ between the events $x_1$ in $R$ and $x^{(r)}$ in $L$. The origin of this reflected point has never been clarified in the literature, including the original paper. 
	
	(b) As we said before, it is well known --- and has been extensively discussed in literature -- that the Minkowski propagator $G_{\mathcal{M}}$ can be thought of as a `thermalised' version of the Rindler propagator $G_{\mathcal{R}}$ in the following sense: 
	\begin{equation}
	G_{\mathcal{M}} (i\tau) = \sum_{n=-\infty}^\infty G_{\mathcal{R}} (i\tau + 2\pi i n)
	\label{thermal}
	\end{equation} 
	This periodicity in the Euclidean time is, of course, the root cause of the thermal behaviour which arises in the standard approach. It is not clear whether the relation in \eq{one} is connected with this basic fact, and if so, how. (We will show that they are intimately related.)
	
	(c) The original derivation makes use of the fact that $G_{\mathcal{R}}$ and $G_{\mathcal{M}}$ are the Feynman propagators in the two vacua. It is not clear whether the same relation holds for a much wider class of functions (and we will see that it does) and if so what are the essential ingredients which go into this relation. 
	(We will discover these ingredients.)
	
	(d) If we analytically continue $\tau \to \tau_E = i\tau$, the coordinate transformation \textit{in the right wedge} changes from $(x=\rho \cosh\tau, t=\rho \sinh \tau)$ to $(x=\rho \cos\tau_E, t_E= \rho \sin\tau_E)$. Therefore, this analytic continuation of the right wedge $alone$ fills the entire Euclidean plane in Cartesian coordinates $(t_E,x)$. In particular, the relation $x= \rho \cosh\tau $ (with $\rho>0$) implies that $x$ is always positive; but on analytic continuation $x=\rho \cos\tau_E$ (with $\rho>0$) covers the entire real line $(-\infty<x<\infty)$ including negative values of $x$.  This, in turn, implies that there is no distinct ``reflected event'' $x^{(r)}$ in the \textit{Euclidean} sector (or, more precisely, both $L$ and $R$ wedges maps to the entire Euclidean plane on their own). So the question arises as to how \eq{one} transforms into the Euclidean sector and how it is to be interpreted. 
	
	(e) The second term in the right hand side of \eq{one} convolves $G_{\mathcal{M}}$ with a kernel of the type $(x^2+\pi^2)^{-1}$. Though the authors of \cite{cr} make a passing comment of some `line charge density', this interpretation does not make any clear physical sense. On the other hand, this particular kernel is well known in complex analysis and is a special case of a Poisson kernel \cite{katznelson2004introduction} with the factor $(x^2 + y^2)^{-1}$ with $y=\pi$. The Poisson kernel can be used to extend functions defined on the real line to the upper half-plane in such a way that they remain bounded (unlike the usual analytic continuation) in the upper half-plane (for example, the standard analytical continuation of 
		$\cos x$ will lead to $\cos z \equiv \cos (x+iy)$ while Poisson kernel will extend $\cos x$ to $e^{-y} \cos x$). Roughly speaking the role of the Poisson kernel is to allow $G_{\mathcal{M}}$ to be extended into the upper complex plane and, in particular, along the line $z=\tau+i\pi$. This is important because the transformation $\tau \to \tau + i\pi$ actually reflects the event  $(x=\rho \cosh\tau, t=\rho \sinh \tau)$ to $(x=-\rho \cosh\tau, t=-\rho \sinh \tau)$ thereby taking $x$ to $x^{(r)}$. Since $G_{\mathcal{M}}$ in the second term of \eq{one} is already evaluated at $x^{(r)}$ , the convolution with a Poisson kernel actually brings it back to $R$. It is not a priori obvious what exactly is going on in the reflection followed by the convolution by the Poisson kernel.
	
	\begin{figure}[h!]
		\centering
		\includegraphics[scale=.3]{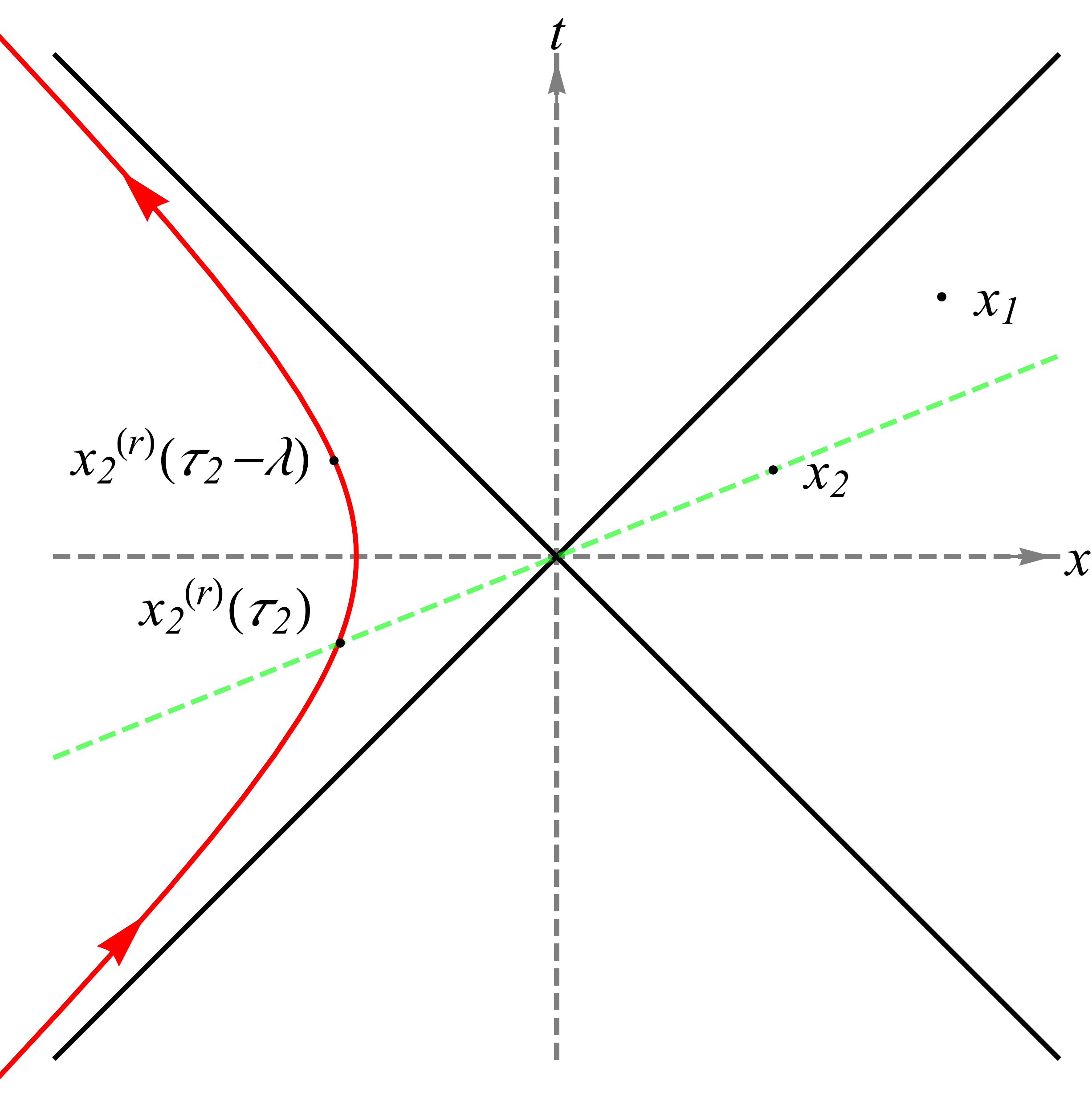}
		\caption{The geometric interpretation of the relation between $x^{(r)}_2$ and $x_{2}$.}
		\label{fig:reflection}
	\end{figure}

	A closely related aspect is the introduction of the transformation $\tau\to\tau+i\pi$ of the Rindler time, which is very different from the usual Euclideanisation obtained by $\tau\to i\tau$. In Euclideanisation we rotate the time axis by $\pi/2$ in going from Re $\tau $ to Im $\tau$. Here we are shifting Re $\tau$ parallel to itself by an amount $\pi$ in the complex plane. The properties of such an analytic extension for various functions play a crucial role in the structure of \eq{one}.

	This paper arose, partially, out of our attempt to probe these issues and to distil the essential mathematical structure behind \eq{one}.  We will show that a large class of functions satisfy \eq{one} --- of which the Feynman propagator is just a special case --- and identify the key structural factor common to all these functions. We will provide a very simple, straightforward proof of \eq{one} for a broad class of functions in Section \ref{genresult} and will follow it up with a more elegant and formal proof which highlights the role of the complex plane, Poisson kernel and the analytic extension $\tau \to \tau+i\pi$.

	\subsection{Motivation 2: Schwinger (heat) kernel and inequivalent vacua}
	
	There is a second motivation for this work which is related to the Schwinger kernel approach to the inequivalent vacua. We will now describe this motivation.
	
	The canonical quantization (based on identifying creation/annihilation operators, vacuum state, Fock basis etc) is the familiar procedure one usually uses. It is, however, possible to approach QFT in curvilinear coordinates/curved spacetime from a different perspective based on the Schwinger (heat) kernel approach. In this approach the central quantity is the Schwinger kernel which, in Lorentzian spacetime, satisfies the covariant differential equation and the boundary condition given by
	\begin{equation}
	i\frac{\partial K}{\partial s} + \Box K = \delta (s) \, \delta(x,x_0); \quad \lim_{s\to 0}K(x,x_0;s) =  \delta (x,x_0)
	\label{pdektp}
	\end{equation}
	where $\delta(x,x_0)$ is the properly densitised Dirac delta function in curved spacetime or curvilinear coordinates. 
	It is possible to obtain all other relevant constructs --- e.g., the Feynman propagator, effective Lagrangian etc. --- from the kernel by:
	\begin{equation}
	iG (x,x_0) = \int_{0}^{\infty}ds K(x,x_0;s) ; \quad  i L_{\rm eff}(x) =\int_{0}^{\infty}\frac{ds}{s} K(x,x_0;s)
	\label{GLtp}
	\end{equation} 
	The \eq{pdektp} shows that $K$ just encodes the properties of the Laplacian in a given curved spacetime and is independent of the physics of the scalar field \cite{note2}.
	There are two aspects related to the Schwinger kernel which we will focus on in this work.
	
	First, the relation between the kernel and the propagator in \eq{GLtp}, as well as the fact that the propagators for two vacuua are related by the thermalisation procedure in \eq{thermaltp} tells us that the two kernels will also satisfy a similar relation:
	\begin{equation}
	K_{\mathcal{M}} (i\tau) = \sum_{n=-\infty}^\infty K_{\mathcal{R}} (i\tau + 2\pi i n)
	\label{thermaltpK}
	\end{equation}
	We will show in Section \ref{inversion} that any two functions which obey this thermalisation condition will also satisfy a result like \eq{one} which can be obtained as the ``inverse'' of the thermalisation condition. Using this fact, we can immediately obtain $K_{\mathcal{R}}$ from the well-known expression for $K_{\mathcal{M}}$. 
	We could not find explicit expressions for the Schwinger kernel, corresponding to Rindler vacuum, in the published literature. 
	Its simple derivation shows the power of our general technique, involved in ``inverting'' \eq{thermaltpK}.

	The above procedure will lead to an expression for $K_{\mathcal{R}}$ which is distinct from $K_{\mathcal{M}}$. This is, of course, as it should be because  
	$K_{\mathcal{R}}$ and $K_{\mathcal{M}}$ have to lead to very different propagators $G_{\mathcal{R}}$ and $G_{\mathcal{M}}$. This fact, however, raises a second question:
	We can incorporate both the differential equation and the boundary condition in \eq{pdektp} and write down the formal solution to this equation in the form 
	\begin{equation}
	K = \exp\left(is\Box\right) \ \delta(x,x_0)
	\label{solk}
	\end{equation} 
	Once we determine $K$ we can obtain the Feynman propagator by using \eq{GLtp}. 
	In this approach, it appears that we are obtaining a Feynman propagator directly from the solution for the Schwinger kernel given in \eq{solk}. It is not a priori obvious how this approach will lead to different vacua, different kernels and different propagators; at first sight, the solution defining the kernel in \eq{solk} appears to be unique and generally covariant, suggesting that there is a unique kernel for the quantum field, independent of the coordinates used to describe it \cite{note3}.
	This, of course, cannot be true because we know from the canonical approach that the definition of propagators also depends on the vacuum state and they are (non-trivially) different when computed using $\ket{\mathcal{M}}$ or $\ket{\mathcal{R}}$. If, instead, we use \eq{solk} to determine $K$ and \eq{GLtp} to determine $G$ the information about vacuum states has to slip in through boundary conditions. It is important to identify how exactly this comes about. 
	
	Interestingly enough, the difference between Rindler and Minkowski vacuum is encoded in the way we choose to represent the Dirac delta function $\delta (x,x_0)$ in \eq{solk}. Notice that in \eq{solk} both the operator $\Box$ as well as $\delta(x,x_0)$ are generally covariant constructs. So if we change the coordinate system, the right-hand side of \eq{solk} will transform in a generally covariant manner and so will the kernel $K(x,x_0; s)$. To get a distinctly different kernel, we need to tinker with the  Dirac delta function.
	Given the rather curious nature of this result, we will discuss it in some detail.  
	(As we said before, we have not seen the expression for $K_{\mathcal{R}}$ in the published literature and the expressions we derive in this paper might also be of intrinsic interest.)
	
	\subsection{Motivation 3: From the Euclidean plane to all the \textit{four} Rindler wedges}\label{sec:mot3}
	
	The standard coordinates in right Rindler wedge $(\tau,\rho)$ are related to the inertial $(t,x)$ coordinates by: $x=\rho\cosh\tau, t=\rho\sinh\tau$ with $\rho>0,-\infty<\tau<\infty$. Clearly the $(\tau,\rho)$ coordinates only cover the right wedge. The analytic continuation $it\to t_E$ and $i\tau\to \tau_E$ leads to $x=\rho\cos\tau_E, t_E=\rho\sin\tau_E$ with  $\rho>0, 0<\tau<2\pi $.
	We see that the Euclidean coordinates $(\tau_E,\rho)$ covers the \textit{entire} Euclidean plane $(t_E,x)$. In other words, the analytic continuation of the Rindler \textit{right wedge alone} fills the entire Euclidean plane, covered by the inertial (Euclidean) coordinates. While the transformation $x=\rho\cosh\tau$ with real $\tau$ only covers $x>0$, the shift $\tau\to\tau+i\pi$ takes us from $x>0$ to $x<0$; the imaginary values of $\tau$ knows about the region beyond the horizon. The horizon ($x^2-t^2=0$) in the Lorentzian sector collapses to the origin ($x^2+t_E^2=0$) in the Euclidean plane and $F,L,P$ wedges disappear. 
	It can be easily checked that the same phenomenon occurs when we analytically continue from, say, the left wedge as well. 
	
	This raises an interesting question, which provides the third motivation for this work. It is generally believed that, at least in flat spacetime, one can define the quantum field theory --- in particular, the Feynman propagator --- properly in the Euclidean sector and then analytically continue it to the Lorentzian sector. This certainly works when we use inertial coordinates and go from $(t_E,x)$ to $(t,x)$. However, it appears that if we start with Rindler (`polar') coordinates $(\tau_E,\rho)$ in the Euclidean sector, write down the Euclidean propagator, and analytically continue to the Rindler coordinates in Lorentzian sector, we only seem to recover the Lorentzian propagator in the right wedge! How do we obtain the form of the propagator when one or both of the coordinates are outside the right wedge? Since the Euclidean Rindler coordinates cover the entire Euclidean inertial manifold, there must exist a way of obtaining the results for all the four wedges in the Lorentzian sector, starting from the Euclidean Rindler coordinates. 
	
	We will see that this is indeed possible, but the analytic continuation is non-trivial, especially when the two events in $G(x_1,x_2)$ are in different sectors. (This was briefly mentioned by one of the authors, in the Appendix of Ref. \cite{Padmanabhan:2019yyg} and we will elaborate on this construction, especially for the Schwinger kernel.) This construction will also provide a relatively simple route to the propagator in different sectors obtained, for example, by a more complicated procedure in
	Ref. \cite{Boulware:1974dm}.  We have not seen the corresponding results for the Schwinger kernel in the literature, and we will provide their explicit forms.

	\section{A general result for a class of functions}\label{genresult}
	
	We will start with the task of understanding \eq{one}. It is well known that the time dependence of the form $\exp(-i\omega|\tau|)$ with $\omega>0$ plays a crucial role in the structure of Feynman propagator. It is this factor which propagates positive frequency modes forward in time and negative frequency modes backward in time.  Consider a function built by an arbitrary superposition of $\exp(-i\omega|\tau|)$ in the form 
	\begin{equation}
	F_{\mathcal{R}}(\tau) = \int_0^\infty d\omega \, A_1(\omega) e^{-i\omega|\tau|} \equiv \int_0^\infty d\omega\, A(\omega) (\sinh \pi \omega) e^{-i\omega|\tau|}
	\label{tpone}
	\end{equation} 
	The first equality defines the superposition in terms of the weightage $A_1(\omega)$; in the second equality we have set $A_1(\omega) \equiv A(\omega) \sinh \pi \omega$ for future convenience. We have only displayed the dependence on the Rindler time coordinate $\tau$ but both $F_{\mathcal{R}}$ as well $A(\omega)$ will depend on all the other coordinates. Most of the time, we will be interested in biscalars which depend on two events $x_1$ and $x_2$  with (i) a dependence in Rindler time coordinates through $\tau \equiv \tau_2-\tau_1$ and (ii) a dependence in the transverse coordinates through $\tv{x} \equiv \tv{x}^2 - \tv{x}^1$. That is, the biscalars respect the translation invariance of the Rindler metric in the Rindler time coordinate and the transverse coordinates.  While dealing with such biscalars, we actually have
	$F_{\mathcal{R}}=F_{\mathcal{R}}(\rho_1, \rho_2, \tv{x},\tau)$ and $A=A(\rho_1, \rho_2, \tv{x},\omega)$ but for clarity, 
	we suppress the display of the dependence on $\rho_1, \rho_2, \tv{x}$ in both $F_{\mathcal{R}}$ and $A(\omega)$. 
	In all these discussions, the transverse coordinates go for ride; so it is often convenient to Fourier transform all the relevant functions with respect to $\tv{x}$ and work with $F_{\mathcal{R}}(\rho_1, \rho_2, \tv{k},\tau)$ and $A(\rho_1, \rho_2, \tv{k},\omega)$; this is what we will do most of the time.
	
	Let us next construct another function obtained by thermalising $F_{\mathcal{R}}(\tau)$ along the lines described just after \eq{thermal}. This is easily done using the result: 
	\begin{equation}
	(\sinh \pi \omega) \sum_{n=-\infty}^\infty e^{-\omega|\tau_E+ 2\pi n|} = \cosh \omega \left( |\tau_E| -\pi\right);\qquad (0<|\tau_E|<2\pi)
	\label{trigsum}
	\end{equation} 
	and analytically continuing back from $\tau_E$ to $\tau$. (For completeness, we have provided a proof of this relation in Appendix \ref{appendixA}). In \eq{trigsum}, the left hand side is clearly periodic in $\tau_E$ with period $2\pi$ by construction. It is obvious that if, say, $\tau_{E}$ lies between $2\pi N$ and $2\pi(N+1)$ the sum can only depend on $\tau_{E}-2\pi N$. In the right hand side, therefore, the range of $\tau_{E}$ is restricted to the interval $(0,2\pi)$. This  procedure leads
	us to the second function $F_{\mathcal{M}}(\tau)$ which is the thermalised version of $F_{\mathcal{R}}(\tau)$:
	\begin{equation}
	F_{\mathcal{M}}(\tau) = \int_0^\infty d\omega\ A(\omega) \cosh \omega \left[ i|\tau| -\pi\right]
	\label{tptwo}
	\end{equation} 
	
	It is now possible to prove that $F_{\mathcal{R}}$ and $F_{\mathcal{M}}$ --- defined by \eq{tpone} and \eq{tptwo} --- satisfy the relation in \eq{one} we are trying to understand, for any choice of $A(\omega)$. In other words, \textit{only two} ingredients have gone into proving the result in \eq{one} for this class of functions \cite{note4}. First is that $F_{\mathcal{R}}$ is built from an arbitrary superpositions of  $\exp(-i\omega|\tau|)$ with $\omega>0$. Second, $F_{\mathcal{M}}$ is constructed by thermalising $F_{\mathcal{R}}$.
	We know that these two conditions are indeed satisfied by the Rindler and Minkowski propagators. The Feynman propagator for the Rindler vacuum is built from a superposition of  $\exp(-i\omega|\tau|)$ with $\omega>0$; and we  know from \eq{thermal} that $G_{\mathcal{M}}$ is a thermalised version of $G_{\mathcal{R}}$. So clearly, these two functions satisfy \eq{one}.
	
	Proving that the functions defined by \eq{tpone} and \eq{tptwo} satisfies the relation like \eq{one} is fairly straightforward. We first note that the factor $A(\omega)$ and the integration over $\omega$ goes for a ride while establishing \eq{one} between $F_{\mathcal{R}}$ and $F_{\mathcal{M}}$. So all we need to prove is an identity satisfied by hyperbolic functions in the form:
	\begin{equation}
	\sinh (\pi \omega) e^{-i\omega|\tau|} = \cosh \omega (i|\tau| -\pi) - \int_{-\infty}^\infty d\mu \ \frac{\cosh \omega [(i|\mu| -\pi) +\pi] }{(\mu -\tau)^2 +\pi^2} 
	\label{finalresult}
	\end{equation}
	(We stress that this is merely an identity involving hyperbolic functions, devoid of physics content.) In the integral in the second term on the right hand side, we actually have just $\cosh (i\omega|\mu|)=\cos\omega\mu$ in the numerator, which we have written in such a manner that  it can be compared with the first term on the right hand side, evaluated with a shift of $\pi$. When $\tau>0$, this shift of $\pi$ in $i\tau$ leads to a shift of $-i\pi$ in $\tau$  and thus ``reflects'' the event from the right wedge to the left wedge. (When $\tau<0$, the shift is by $i\pi$, which again leads to the same `reflection'.) Once we have the result in \eq{finalresult} we can multiply the whole equation by $A(\omega)$ and integrate over $\omega$ to establish that $F_{\mathcal{M}}$ and $F_{\mathcal{R}}$ satisfy  the relation in \eq{one}. 
	
	The proof of \eq{finalresult} is straightforward, almost trivial. Writing the numerator of the integral as $\cosh (i\omega|\mu|)=\cos\omega\mu$ and performing the integral, we get this term to be:
	\begin{equation}
	- \int_{-\infty}^\infty d\mu\ \frac{\cos \omega \mu}{(\mu - \tau)^2 + \pi^2} =- e^{-\pi \omega} \cos \omega \tau
	\label{note3}
	\end{equation}
	The result in  \eq{finalresult} now reduces to a simple, easily verified,  hyperbolic identity:
	\begin{equation}
	(\sinh \pi \omega) \, e^{-\omega z} = \cosh \omega [z-\pi] - e^{-\pi \omega} \cosh \omega z,
	\label{note4}
	\end{equation}
	evaluated for   $z = i|\tau|$.  This completes the proof of \eq{finalresult}. As explained earlier, the only two ingredients which went into establishing this result are: (i) The $F_{\mathcal{R}}$ is built by a superposition of $\exp-i\omega|\tau|$ and (ii) $F_{\mathcal{M}}$ is obtained from $F_{\mathcal{R}}$ by thermalisation.

	It is easy to verify that the Rindler and Minkowski propagators indeed have the form of $F_{\mathcal{M}}$ and $F_{\mathcal{R}}$ with a particular choice for $A(\omega)$ given by 
	\begin{equation}
	A(\omega) = \frac{1}{\pi^2} K_{i\omega}(\mu \rho_1) K_{i\omega}(\mu \rho_2) ; \qquad \mu^2 = m^2 + \tv{k}^2
	\label{choiceofA}
	\end{equation} 
	With this choice of $A(\omega)$ in \eq{tpone} and \eq{tptwo} we will correctly reproduce the two propagators $G_{\mathcal{R}}(\tau,\tv{k},\rho_1,\rho_2)$ and $G_{\mathcal{M}}(\tau,\tv{k},\rho_1,\rho_2)$
	which are the Fourier transforms of the propagators in transverse coordinates.  The cognoscenti will immediately see that \eq{tpone}, with the $A(\omega) $ in \eq{choiceofA}, gives the Rindler propagator because it is built from the normalized mode functions of the form $u_{\omega \tv{k}} \propto (\sinh \pi \omega)^{1/2} K_{i\omega} (\mu \rho) \, e^{i(\tv{k} \cdot \tv{x} - \omega\tau)}$. Proving that \eq{choiceofA}, substituted into \eq{tptwo} gives the Minkowski propagator is also not difficult. To do this most directly, start from the Schwinger representation for the Minkowski propagator given by 
	\begin{equation}
	G_{\mathcal{M}}(x_1,x_2) = i \frab{1}{4\pi i}^{D/2} \int_0^\infty \frac{d\lambda}{\lambda^{D/2}} \ e^{-i\lambda m^2 - (i/4\lambda)\sigma^2}
	\end{equation} 
	where $\sigma^2(x_1,x_2)$ is the interval between the two events in $R$ expressed in Rindler coordinates. It contains the square of transverse separation $(\Delta \tv{x})^2$. Fourier transforming with respect to this separation leads to the result 
	\begin{equation}
	G_{\mathcal{M}}(\tau, \tv{k}, \rho_1,\rho_2) = \frac{1}{2\pi} K_0 (\mu \ell)
	\end{equation} 
	where $\ell^2 \equiv \rho_1^2 + \rho_2^2 - 2\rho_1\rho_2 \cosh\tau $ and $\mu^2 =m^2+\tv{k}$.  We now recall the convenient identity (see Eq. (24), p382 of \cite{bateman1954tables})
	\begin{equation}
	\frac{\pi}{2}  K_0 (\mu \ell) = \int_0^\infty d\omega \, K_{i\omega}(\mu \rho_1) K_{i\omega}(\mu \rho_2) \cosh [\omega(\pi - \tau_E)]
	\end{equation} 
	analytically continue it from $\tau_E$ to $\tau$, and express $K_0(\mu\ell)$ as a product of two $K_{i\omega}$ functions. This will prove that the choice of $A(\omega)$ in \eq{choiceofA}, when substituted into \eq{tptwo}, does reproduce the Minkowski propagator. 
	
	We will now comment on a wider class of functions for which this result holds. Any even function, $F_{\mathcal{R}}(\tau)$, can be expressed in the form:
	\begin{equation}
	F_{\mathcal{R}}(\tau)=\int_{-\infty}^{+\infty}d\omega B(|\omega|)\exp (-i\omega|\tau|)
	\label{more-general}
	\end{equation} 
	which differs from our original set in \eq{tpone} by the fact that we are now integrating over the whole range of $\omega$ without just restricting to the positive frequencies. 
	Once we obtain the result in \eq{finalresult} we can multiply the whole equation by  $B(|\omega|)$ and integrate overall $\omega$ and again obtain the result in \eq{one} for all $F_{\mathcal{R}}(\tau)$ which is even in $\tau$. 
	A simple example of such a function is the Schwinger (heat) kernel defined with respect to Rindler and Minkowski vacua. The Rindler heat kernel $K_{\mathcal{R}}(\tau)$ is an even function of $\tau$ and can indeed be expressed in the form an integral in \eq{more-general}. So the result in \eq{one} holds not only for the propagators but also for the heat kernels and we will discuss these features in detail in Sections \ref{kernelforR} and \ref{heatkernel}.  
	
	We will now provide a more elegant proof of \eq{one} for all even functions $F_{\mathcal{R}}(\tau)$ which clearly proves that it is an ``inversion'' of the thermalisation condition.

	\section{Inversion of the thermalisation condition: An elegant approach}\label{inversion}
	
	We will next provide a more general, but somewhat formal proof of the results obtained in the last section. This proof has the advantage that --- besides being fairly elegant --- it can be generalised to a wide class of spacetimes and a variety of functions related to Rindler-like and Minkowski-like vacua. The basic procedure is as follows.
	We start with two functions $F_{\mathcal{M}}(\tau)$ and $F_{\mathcal{R}}(\tau)$ related by the thermalisation condition in \eq{thermal}.
	The only assumption we make is that $F_{\mathcal{R}}(z) = F_{\mathcal{R}}(-z)$, viz. $F_{\mathcal{R}}$ is an even function of its relevant argument.
	We then express the thermalisation condition in the form of an integral relation given by
	\begin{equation}
	F_{\mathcal{M}}(z)=\int_{\mathcal{C}'}\frac{du}{(i\pi)}\mathcal{H}(z;u)F_{\mathcal{R}}(u).
	\label{k58}
	\end{equation} 
	where $\mathcal{H}$ is a suitable integral kernel and $\mathcal{C}'$ is a specific contour in the complex plane given in Figure 3(a). We have implicitly assumed that $F_{\mathcal{R}}$ has the required limits for the right hand side of \eq{k58} to exist. Once $F_{\mathcal{M}}$ is related to $F_{\mathcal{R}}$ by this integral equation, we can invert the kernel and find $F_{\mathcal{R}}$ in terms of $F_{\mathcal{M}}$ thereby again getting a relation of the form 
	\begin{equation}
	F_{\mathcal{R}}(z)=\int_{\mathcal{C}}\frac{du}{(i\pi)}\mathcal{G}(z;u)F_{\mathcal{M}}(u)
	\label{k65}
	\end{equation} 
	where $\mathcal{G}$, whose explicit form will be given later, is the `inverse' of integral kernel $\mathcal{H}$ which appears in \eq{k58} and we shall soon see that $\mathcal{C}$ is given by the contour in Figure 3(b). (Again, we require that $F_{\mathcal{M}}$ vanishes sufficiently faster on the asymptotic regions of $\mathcal{C}$ for the last integral to exist.) This result, in turn, will lead to an expression of the form in \eq{one} for the two functions $F_{\mathcal{M}}$ and $F_{\mathcal{R}}$. We will now provide the details of this approach.

	Since $F_{\mathcal{M}}$ is an infinite periodic sum of $F_{\mathcal{R}}$, \eq{k58} will lead to a result like \eq{thermal} if $\mathcal{H}$ has poles at 
	\begin{equation}
	u=\pm z+2\pi i n; \qquad n\in \mathbb{Z}
	\end{equation} 
	with a constant value of residues, independent of $n$. There are several functions which will satisfy this criterion and we will choose for our purpose the function 
	\begin{equation}
	\mathcal{H}(z;u)=\frac{\sinh u}{4(\cosh u-\cosh z)}
	\label{k59}
	\end{equation} 
	with the integration contour $\mathcal{C}'$ shown in Figure 3(a) and $p_i$ (see the figure) represents a generic pole of the function $F_{\mathcal{R}}$.  A straightforward calculation using the residue theorem now leads to the result
	\begin{align}
	F_{\mathcal{M}}(z)&=\int_{\mathcal{C}'}\frac{du}{(i\pi)}\mathcal{H}(z;u)F_{\mathcal{R}}(u)\\\nonumber
	&=\frac{1}{2}\sum_{n=-\infty}^{\infty}F_{\mathcal{R}}(z+2\pi i n)+\frac{1}{2}\sum_{n=-\infty}^{\infty}F_{\mathcal{R}}(-z+2\pi i n)\\\nonumber
	&=\sum_{n=-\infty}^{\infty}F_{\mathcal{R}}(z+2\pi i n)
	\end{align}
	where the last step uses our assumption that $F_{\mathcal{R}}(z)$ is an even function. This completes the first part of our task, viz., expressing $F_{\mathcal{M}}$ as a contour integral involving $F_{\mathcal{R}}$. 
	\begin{figure}[h!]
		\centering
		\begin{subfigure}[b]{0.4\linewidth}
			\includegraphics[scale=.25]{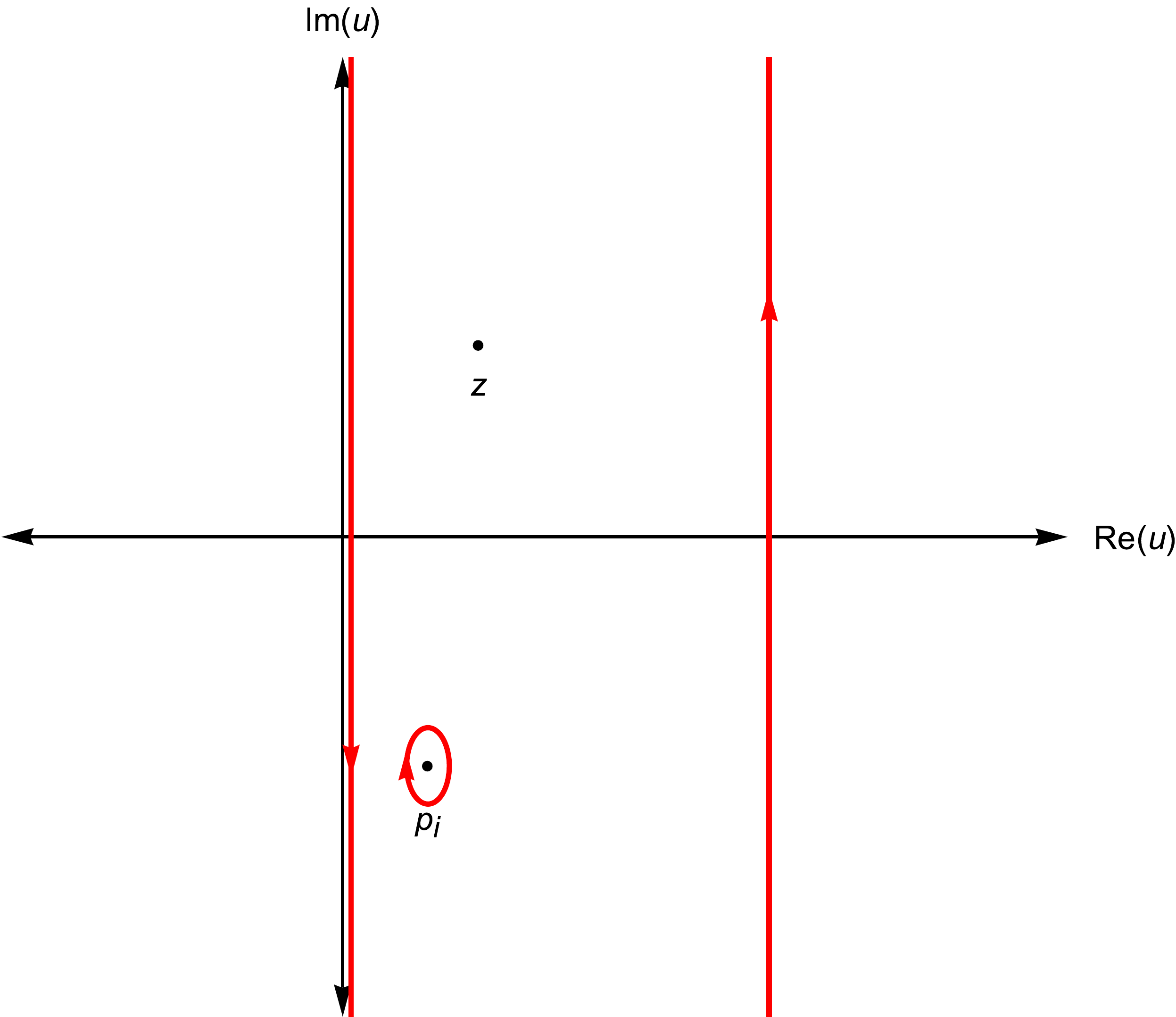}
			\caption{The contour $\mathcal{C}'$ used in \eq{k58}}
			\label{fig:2}
		\end{subfigure}
	\hspace{.5cm}
		\centering
		\begin{subfigure}[b]{0.5\linewidth}
			\includegraphics[scale=.25]{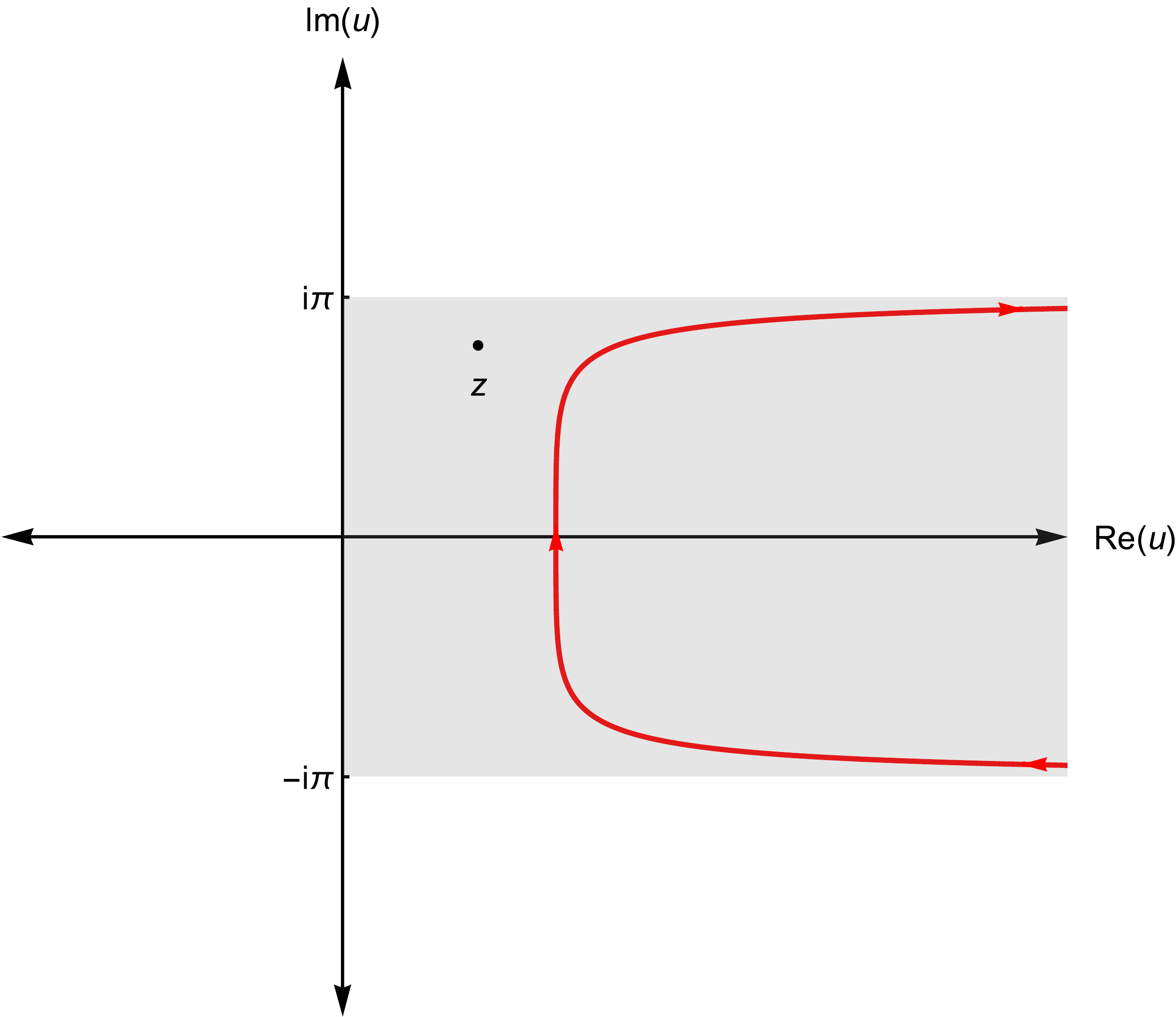}
			\caption{The contour $\mathcal{C}$ used in \eq{k65} and \eq{defInu}}
			\label{fig:1}
		\end{subfigure}
		\caption{The two contours relevant for the `inversion' of thermalisation condition }
		\label{fig}
	\end{figure} 
	The next step is to invert the relation in \eq{k58} and express $F_{\mathcal{R}}$ as an integral over $F_{\mathcal{M}}$ in the form of \eq{k65}. It turns out that the relevant inverse function to be used in \eq{k65} is given by
	\begin{equation}
	\mathcal{G}(z;u)=\frac{u}{(u^2-z^2)}
	\label{k66}
	\end{equation} 
	where the contour is $\mathcal{C}$ shown in Figure 3(b). We will first demonstrate that, with this choice of $\mathcal{G}$ and the contour $\mathcal{C}$, we do reproduce a relation of the form in \eq{one}. Having done that, we will provide a direct proof that $\mathcal{G}$ and $\mathcal{H}$ are indeed inverses of each other in Appendix \ref{AppendixB}. 
	
	\begin{figure}
		\centering
		\includegraphics[scale=.3]{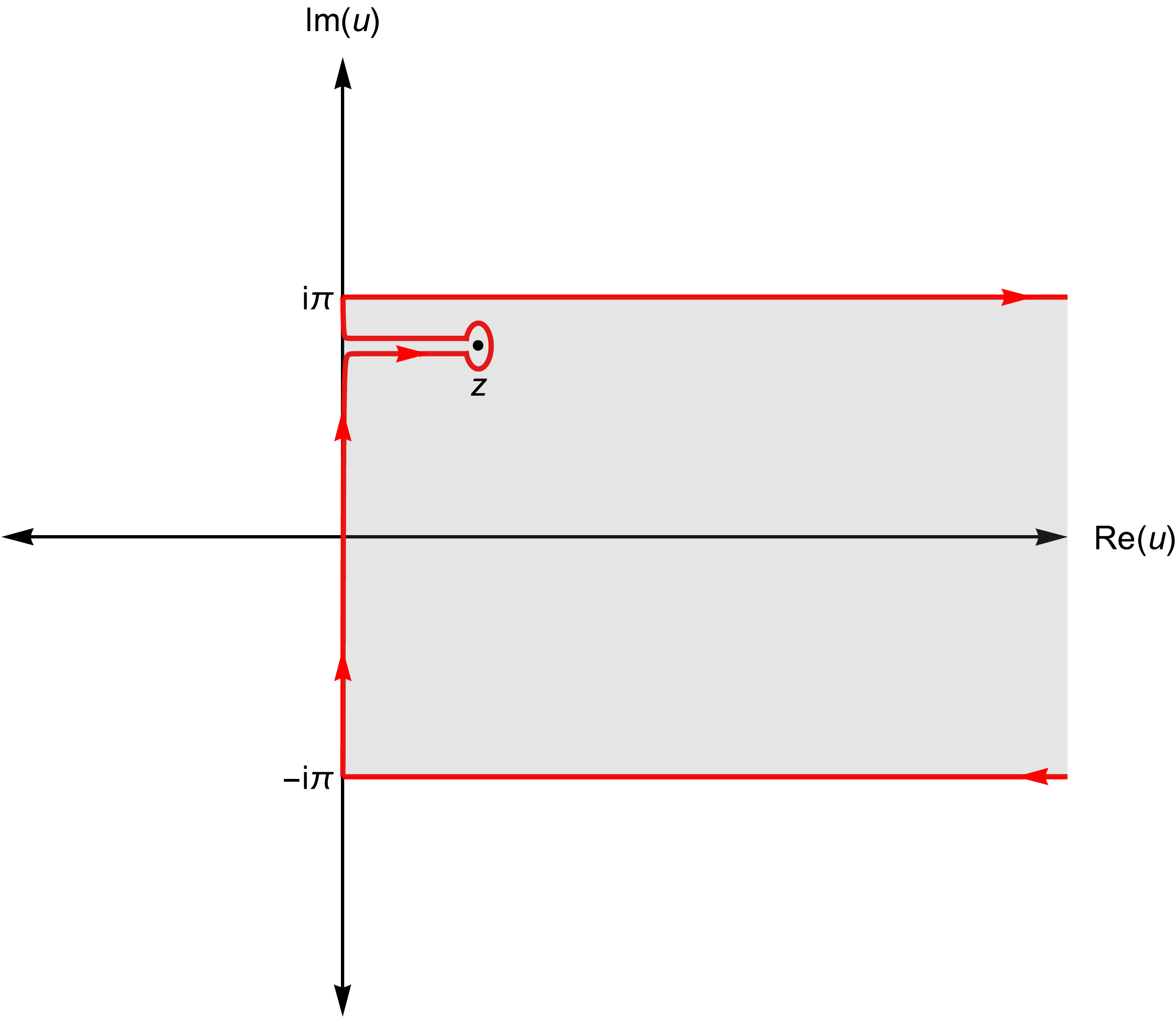}
		\caption{The contour $\tilde{\mathcal{C}}$, which is obtained by deforming $\mathcal{C}$.}
		\label{fig2_2}
	\end{figure}
	
	We shall now assume that the function $F_{\mathcal{M}}$ is analytic in the shaded region in Figure 3(b). Therefore, we can deform the contour $\mathcal{C}$ into $\tilde{\mathcal{C}}$ as shown in Figure 4. The integral in \eq{k65}, evaluated along the deformed contour $\tilde{\mathcal{C}}$, can then be written conveniently as the sum of three integrals as follows:   
	\begin{align}
	\int_{\mathcal{C}}\frac{du}{(i\pi)}\mathcal{G}(z;u)F_{\mathcal{M}}(u)&=\textrm{Residue term + vertical integral + horizontal integrals}
	\end{align}
	The reside term is just $F_{\mathcal{M}}(z)$. Since $F_{\mathcal{M}}(z)$ is an even function,  $\mathcal{G}(z;iy)F_{\mathcal{M}}(iy)$ is an odd function of $y$, for real $y$. Hence, the vertical integral vanishes. Let us now look at the horizontal integrals.
	\begin{align}
	\textrm{Horizontal integrals}=\frac{i}{\pi}\int_{\infty}^{0}\frac{(\lambda-i\pi)}{(\lambda-i\pi)^2+z^2}F_{\mathcal{M}}(\lambda-i\pi)d\lambda\,+\,\frac{i}{\pi}\int_{0}^{\infty}\frac{(\lambda+i\pi)}{(\lambda+i\pi)^2+z^2}F_{\mathcal{M}}(\lambda+i\pi)d\lambda
	\end{align}
	The evenness and pseudo-periodicity condition of $F_{\mathcal{M}}$ implies that $F_{\mathcal{M}}(\lambda-i\pi)=F_{\mathcal{M}}(\lambda+i\pi)$.
	Using this fact, followed by some simplifications, we arrive at:
	\begin{align}
	\textrm{Horizontal integrals}
	&=-\int_{-\infty}^{\infty}d\lambda\ \frac{F_{\mathcal{M}}(\lambda+i\pi)}{\pi^2+(\lambda-z)^2}
	\end{align}
	This gives us the final result
	\begin{equation}
	F_{\mathcal{R}}(z)=F_{\mathcal{M}}(z)-\int_{-\infty}^{\infty}d\lambda\ \frac{F_{\mathcal{M}}(\lambda+i\pi)}{\pi^2+(\lambda-z)^2}
	\label{k72}
	\end{equation} 
	which has the same form as in \eq{one} whenever the transformation $\lambda\rightarrow\lambda+i\pi$ leads to the `reflection' that we discussed in Section \ref{mot1} (also, see Figure 2). Thus the result in \eq{one} is quit general and holds under the following generic conditions: (a) $F_{\mathcal{R}}(z)$ is an even function of $z$ and $F_{\mathcal{M}}(z)$ is obtained by the thermalisation of the function $F_{\mathcal{M}}$. We do not use any other specific property of the spacetime or the nature of the two vacua etc. 
	
	The very general nature of our proof allows it to be extended to several other spacetimes, like, e.g., Schwarzschild, deSitter etc. Whenever the spacetime has a bifurcate Killing horizon, the orbits of the Killing vector allow the partitioning of the relevant plane into four wedges just as in the case of Rindler coordinatisation of Minkowski spacetime. In a general context, we will not have explicit/closed-form expressions for the mode functions of the scalar field, and hence one may not be able to carry out explicit computation of, say, the propagators $G_{\mathcal{R}}$ and $G_{\mathcal{M}}$. However, in all these contexts (e.g., Schwarzschild, deSitter, etc.) we can prove that the propagator in the global, Minkowski-like vacuum, $G_{\mathcal{M}}$ is a thermalised version of the propagator $G_{\mathcal{R}}$ defined in the Rindler-like vacuum.  Given this thermalisation condition and the evenness of $G_{\mathcal{R}}(\tau)$ one can immediately obtain \eq{k72}  in all these spacetimes. To complete the generalisation, we should be able to arrange matters such that the shift 
	$\tau+i\pi$ leads to the reflected trajectory. This holds, again, in any spacetime with a bifurcate Killing horizon. 
	In such spacetimes, 
	one can introduce coordinate systems in such a way that the thermalisation condition --- viz., the periodicity of $\tau_E$ in $2\pi$ --- arises as a two-step process; the shift $\tau_E\to \tau_E + i\pi$ reflects the coordinates from the right wedge to the left wedge and a further shift by $-i\pi$ brings it back to the right wedge thereby ensuring periodicity.
	So our result holds in all these contexts which is a \textit{significant generalisation} of the original result of Ref. \cite{cr}.
	
	All that remains to be shown is that the integral kernels $\mathcal{G}$ and $\mathcal{H}$ (appearing in \eq{k58} and \eq{k65})  given by \eq{k59} and \eq{k66} are indeed inverses of each other. To do this, we only need to show that \eq{k65} implies \eq{k58} and vice-versa with these choices. This is completely straightforward, and hence we delegate the technical details to Appendix \ref{AppendixB}.

	The result in \eq{k72} implies a simple relationship between the Fourier transform of $F_{\mathcal{M}}$ and $F_{\mathcal{R}}$. To derive this relation, we can restrict ourselves to real values of $z$. For simplifying the notation, we will write  $F_{\mathcal{M}}(\lambda+i\pi)\equiv F_{\mathcal{M}}^{(r)}(\lambda)$ with the superscript ``r'' reminding us of the `reflection' in the coordinates. Then, for real values of $z$, our result in \eq{k72} reduces to: 
	\begin{equation}
	F_{\mathcal{R}}(\tau)=F_{\mathcal{M}}(\tau)-\int_{-\infty}^{\infty}d\lambda\ \frac{F_{\mathcal{M}}^{(r)}(\lambda)}{\pi^2+(\lambda-\tau)^2};\qquad (\tau\in\mathbb{R})
	\label{k72dash}
	\end{equation}
	We will now take the Fourier transform of both sides of this equation with respect to $\tau$. 
	Note that, the second term in \eq{k72dash} is just a convolution of $F_{\mathcal{M}}^{(r)}$ with a normalized Lorentzian function, which has the  Fourier transform  $e^{-\pi|\omega|}$. So  the convolution theorem leads to the final result:
	\begin{align}\label{ftrelation}
	\tilde{F}_{\mathcal{R}}(\omega)=\tilde{F}_{\mathcal{M}}(\omega)-e^{-\pi|\omega|}\tilde{F}_{\mathcal{M}}^{(r)}(\omega)
	\end{align}
	where the tilde over a function indicates the Fourier transform (see \cite{note5} for a cautionary note). 
	We will see later that, this remarkably simple relation can be used to derive the Fourier transform with respect to the Rindler time $\tau$ of $G_{\mathcal{R}}$, the Feynman propagator in the Rindler vacuum. Obtaining the same by direct calculation involves a tricky deformation of a contour integral. Further, by choosing $F_{\mathcal{R}}$ and $F_{\mathcal{M}}$ appropriately, the relation \eq{ftrelation} can be applied to Feynman propagators in the appropriate vacuum states in \textit{any} (bifurcate Killing) horizon in curved spacetime.

	\begin{figure}
		\centering
		\includegraphics[scale=.3]{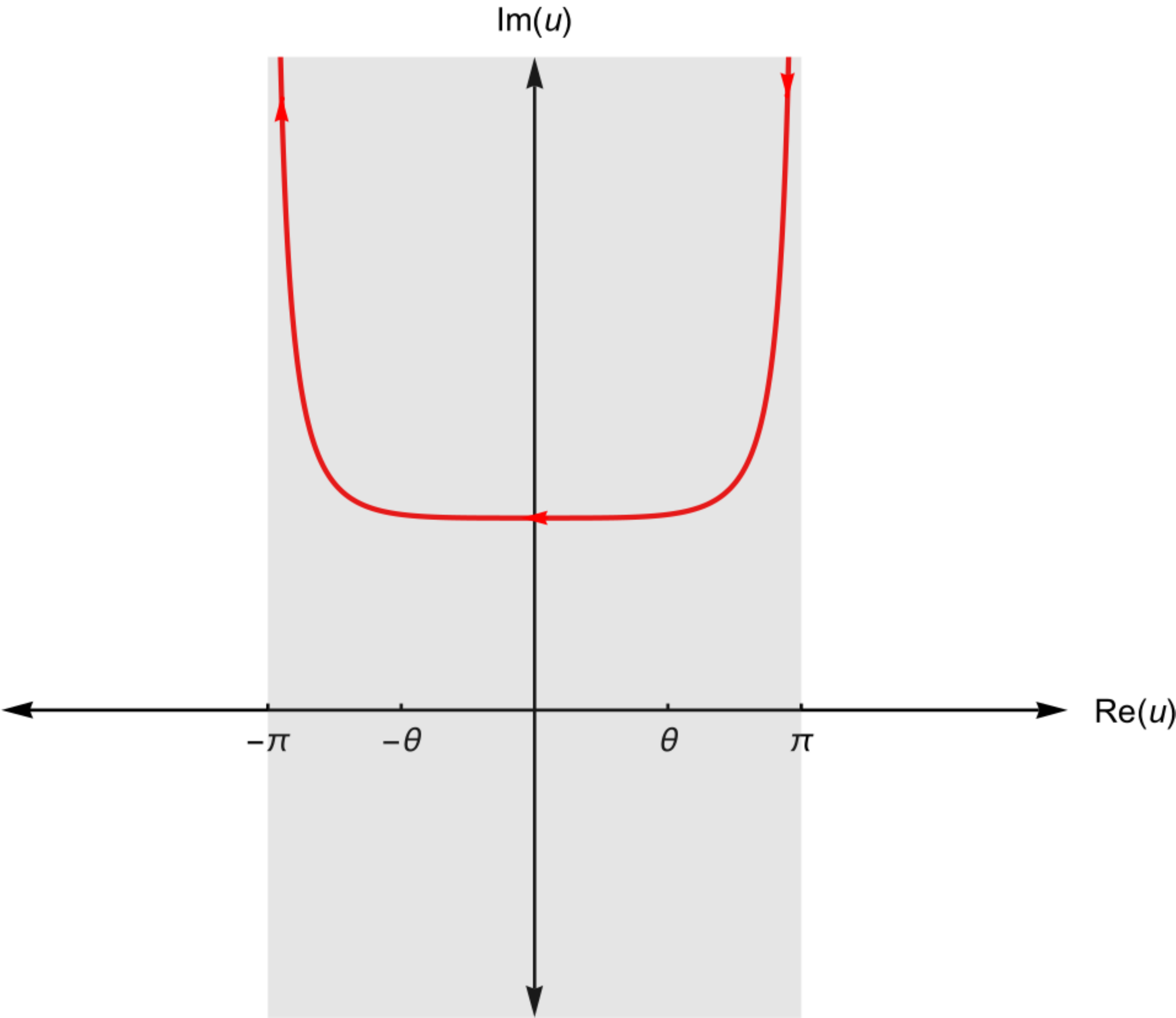}
		\caption{The contour of integration in \eq{Euclidinvert}.}
		\label{fig3}
	\end{figure}

	Before concluding this section, we will briefly consider the Euclidean version of our results. 
	For this purpose, let us  consider an even function $f_{\mathcal{M}}(z)$ which satisfies the condition $f_{\mathcal{M}}(z+2\pi n)=f_{\mathcal{M}}(z)$. Hence, along real line, $f_{\mathcal{M}}$ is a periodic function with a period $2\pi$. Let us now suppose that we can write $f_{\mathcal{M}}(z)$ as a periodic sum of another even function $f_{\mathcal{R}}(z)$. That is,
	\begin{align}
	f_{\mathcal{M}}(z)=\sum_{n=-\infty}^{\infty}f_{\mathcal{R}}(z+2\pi n)
	\end{align} 
	A natural question to ask is: can we retrieve the function $f_{\mathcal{R}}(z)$ from $f_{\mathcal{M}}(z)$ by an integral transformation analogous to \eq{k65}. By an argument similar to the previous one in the Lorentzian sector, we can see that this transformation is given by
	\begin{align}\label{Euclidinvert}
	f_{\mathcal{R}}(z)=\int_{\pi+i\infty}^{-\pi+i\infty}f_{\mathcal{M}}(u)\left(\frac{2u}{u^2-z^2}\right)\frac{du}{(2\pi i)}
	\end{align} 
	where, the integration is along the contour shown in Figure 5. Let us now restrict the variable $z$ to be a real number in the range $(-\pi,\pi)$ (which can be done without loss of generality, when $z$ is real), and call the variable $z$ with this restriction as $\theta$. When $f_{\mathcal{M}}$ is analytic in the shaded region of Figure 5, we can deform the contour such that the integral in \eq{Euclidinvert} can be broken down into four parts: (i) the integral along the vertical line joining $\pi+i\infty$ and $\pi$ (ii) the integral along the vertical line joining $-\pi$ and $-\pi+i\infty$ (ii) integral from $\pi$ to $-\pi$ along the real line and (iv) the integral along infinitesimal semicircles around the poles of the integrand at $z=\pm\theta$. This is explicitly given below:
	\begin{align}
	f_{\mathcal{R}}(\theta)&=\int_{\infty}^{0}\frac{2(\pi+i\lambda)f_{\mathcal{M}}(\pi+i\lambda)}{(\pi+i\lambda)^2-\theta^2}\frac{d\lambda}{(2\pi)}+\int_{0}^{\infty}\frac{2(-\pi+i\lambda)f_{\mathcal{M}}(-\pi+i\lambda)}{(-\pi+i\lambda)^2-\theta^2}\frac{d\lambda}{(2\pi)}\\\nonumber
	&+\int_{\pi}^{-\pi}\frac{2xf_{\mathcal{M}}(x)}{x^2-\theta^2}dx+
	\frac{1}{2}\textrm{Res}\left[f_{\mathcal{M}}(u)\left(\frac{2u}{u^2-\theta^2}\right)\right]_{u=\theta}+\frac{1}{2}\textrm{Res}\left[f_{\mathcal{M}}(u)\left(\frac{2u}{u^2-\theta^2}\right)\right]_{u=-\theta}
	\end{align}
	The last two terms combine to give $f_{\mathcal{R}}(\theta)$. The third term vanishes because the integrand is an odd function. The surviving terms can be combined to get the following expression for $f_{\mathcal{R}}(\theta)$:
	\begin{align}
	f_{\mathcal{R}}(\theta)=f_{\mathcal{M}}(\theta)-\int_{-\infty}^{\infty}\frac{f_{\mathcal{M}}(\pi+i\lambda)}{\pi^2+(\lambda-i\theta)^2}d\lambda;\qquad|\theta|<\pi
	\end{align}
	Once again, we see that the Poisson-like kernel emerges. Note that this is precisely the form that one expects, given our result in \eq{k72}. 
	
	\section{The Schwinger kernel for the Rindler vacuum}\label{kernelforR}
	
	We shall now turn our attention to the derivation and discussion of the Schwinger kernel corresponding to the Rindler vacuum. In particular, we will concentrate on two aspects. First, we will use the relations obtained in the previous sections, which allows us to invert the thermalisation property and obtain the Rindler kernel in the Lorentzian sector directly. This shows the power of the result obtained in Section \ref{inversion} by its application in a new non-trivial context. Second, we will address the question of how the Rindler kernel can be obtained in the Euclidean sector. This requires careful consideration of the boundary conditions, and we will provide a derivation from first principles working entirely in the Euclidean sector. We will start by inverting the thermalisation condition in the Lorentzian sector. 
	
	\subsection{Inversion of the thermalisation condition}

	As we said before, the Schwinger kernel also satisfies the thermalisation condition in the form of \eq{thermaltpK}. Therefore, our result in Section \ref{inversion} tells us that it can be obtained from the Minkowski kernel through the relation
	\begin{equation}
	K_{\mathcal{R}}(\tilde{\tau})=\int_{\mathcal{C}}\frac{du}{(i\pi)}\mathcal{G}(\tilde{\tau};u)K_{\mathcal{M}}(u)
	\label{k78}
	\end{equation} 
	where the ``inverter'' is given by
	\begin{equation}
	\mathcal{G}(\tilde{\tau};u)=\frac{u}{u^2-\tilde{\tau}^2}
	\end{equation} 
	and the Minkowski kernel, expressed in Rindler coordinates, is given by: 
	\begin{equation}
	K_{\mathcal{M}}=\frac{1}{(4\pi i s )}\exp\left(-\frac{\rho'^2+\rho^2}{4i s }\right) \exp\left(z \cosh \tilde{\tau}\right)\equiv
	\mathcal{N}(\rho,\rho';s)\exp\left(z \cosh \tilde{\tau}\right)
	\end{equation} 
	with 
	\begin{equation}
	z=\frac{\rho\rho'}{2i s } ; \qquad\qquad \tilde{\tau}=|\tau-\tau'|
	\end{equation}
	Note that `$s$' actually stands for $|s|\, e^{-i\epsilon}$ to ensure that integrals over $s$ converge (we have, however, not explicitly displayed the phase factor $e^{-i\epsilon}$ almost anywhere in this work). We have also not bothered to include the transverse part and the mass term, since they go for a ride during the ``inversion'' process. All the non-trivial aspects of the kernels that we need in this section are in their $\tau,\rho$ dependent parts. Therefore, for the purpose of deriving the main result of this section, it suffices to work in the 1+1 dimensions. However, for completeness, by suitable generalization of 1+1 dimensional results, we shall explicitly present the d+1 dimensional case, towards the end.

	In order to evaluate the integral in \eq{k78} we shall first assume that Im $\bar\tau < -\pi$ and then analytically continue our result to Im $\bar\tau =0$. We can then use 
	\begin{align}
	\frac{u}{u^2-\tilde{\tau}^2}=\int_{0}^{\infty}d\omega\sin(\omega u)e^{-i\omega \tilde{\tau}};\quad(\textrm{Im}[\tilde{\tau}]<-|\textrm{Im}[u]|)
	\end{align}
	to rewrite $K_{\mathcal{R}}$ as,
	\begin{align}
	K_{\mathcal{R}}(\tilde{\tau})=\mathcal{N}\int_{\mathcal{C}}\frac{du}{(i\pi)}\left(\int_{0}^{\infty}d\omega\sin(\omega u)e^{-i\omega \tilde{\tau}}\right)e^{z\cosh u};\quad(\textrm{Im}[\tilde{\tau}]<-|\textrm{Im}[u]|)
	\end{align}
	The order of integration can now be interchanged to get,
	\begin{align}
	K_{\mathcal{R}}(\tilde{\tau})&=\mathcal{N}\int_{0}^{\infty}d\omega e^{-i\omega \tilde{\tau}}\left(\int_{\mathcal{C}}\frac{du}{(i\pi)}\sin(\omega u)e^{z\cosh u}\right);\quad(\textrm{Im}[\tilde{\tau}]<-|\textrm{Im}[u]|)    
	\end{align}
	Recall that the Modified Bessel function $I_{\nu}$ has the following integral representation \cite{watson1995treatise}.
	\begin{align}\label{defInu}
	I_{\nu}(z)&=\int_{\mathcal{C}}\frac{du}{2\pi i}e^{z\cosh u}e^{-\nu u}
	\end{align}
	where, the contour $\mathcal{C}$ is as shown in Figure 3(b). Hence,  the expression for $K_{\mathcal{R}}$ when $\textrm{Im}[\tilde{\tau}]<-\pi$ reduces to:
	\begin{align}                                 
	K_{\mathcal{R}}(\tilde{\tau})&=\mathcal{N}\int_{0}^{\infty}\frac{d\omega}{i} e^{-i\omega\tilde{\tau}}\left[I_{-i\omega}(z)-I_{i\omega}(z)\right]\\
	&=\frac{1}{(2\pi i s )}\exp\left(-\frac{\rho^2+\rho'^2}{4i s }\right) \int_{0}^{\infty}\frac{d\nu}{\pi} e^{-i\nu \tilde{\tau}}\sinh(\pi\nu)K_{i\nu}\left(\frac{\rho \rho'}{2i s }\right)
	\label{k86}
	\end{align}
	%%%%
	%\begin{figure}
	%	\centering
	%	\includegraphics[scale=.3]{plot0.pdf}
	%	\caption{The contour $\mathcal{C}$.}
	%	\label{fig1}
	%\end{figure}
	%%%
	The Rindler kernel for $\tilde{\tau}>0$, in principle, can now be defined as the analytic continuation of $K_{\mathcal{R}}(\tilde{\tau})$ to positive real axis in the complex $\tilde{\tau}$-plane.   Unfortunately, the final integral in \eq{k86} does not seem to have a closed expression (in terms of standard functions) that we can use to analytically continue $K_{\mathcal{R}}(\tilde{\tau})$ for positive values of $\tilde{\tau}$ in a straightforward manner. However, by computing the corresponding propagator from $K_{\mathcal{R}}(\tilde{\tau})$, we shall shortly
	verify that it is in fact the correct Rindler kernel.     
	
	We now provide the result for the, massive, $d+1$ dimensional case. Let $K^{(m)}_{\mathcal{R}}(\mathbf{k}_{\perp},\tilde{\tau})$ be the Fourier transform  of the massive kernel $K^{(m)}_{\mathcal{R}}$ with respect to the transverse Cartesian coordinate differences $\Delta\mathbf{x}_{\perp}$. We also recall that the massive kernel is just $e^{-im^2s}$ times the massless kernel. Hence, the generalization of Eq.(\ref{k86}) can be conveniently written in terms of $K^{(m)}_{\mathcal{R}}(\mathbf{k}_{\perp},\tilde{\tau})$ as,
	\begin{align}
	K^{(m)}_{\mathcal{R}}(\mathbf{k}_{\perp},\tilde{\tau})&=\frac{\exp\left(-is\mu^2\right)}{(2\pi i s )}\exp\left(-\frac{\rho^2+\rho'^2}{4i s }\right) \int_{0}^{\infty}\frac{d\nu}{\pi} e^{-i\nu \tilde{\tau}}\sinh(\pi\nu)K_{i\nu}\left(\frac{\rho \rho'}{2i s }\right) ;\qquad \textrm{Im}[\tilde{\tau}]<-\pi.
	\label{k86gen}
	\end{align} 
	where, $\mu^2=m^2+|\mathbf{k}|^2$. (The $\mathbf{k}_{\perp}$-dependent exponential factor coming from the Fourier transform of $\exp\left[-\Delta\mathbf{x}_\perp^2/(4is)\right]$ with respect to $\Delta\mathbf{x}_{\perp}$ and $\exp\left(-im^2s\right)$ combine to give the $\mu^2$ dependent part of the kernel. This is why working with the Fourier transform with respect to the transverse coordinates is useful.)  The rest of the terms in \eq{k86gen} can be seen to match exactly with the 1+1 dimensional massless kernel given in Eq.(\ref{k86}). To check the correctness of \eq{k86}, let us calculate the Feynman propagator $G_{\mathcal{R}}$ for a massive scalar field from $K^{(m)}_{\mathcal{R}}(\mathbf{k}_{\perp},\tilde{\tau})$ and compare it with the known results in literature. The Fourier transform $G_{\mathcal{R}}(\mathbf{k}_{\perp},\tilde{\tau})$ with respect to $\mathbf{x}_{\perp}$ of the Feynman propagator $G_{\mathcal{R}}$ for a massive scalar field (when $\textrm{Im}[\tilde{\tau}]<-\pi$) can be evaluated using \eq{GLtp} to give:
	\begin{align}\label{GRexp}
	G_{\mathcal{R}}(\mathbf{k}_{\perp},\tilde{\tau})=-i\int_{0}^{\infty}ds\left[\frac{\exp\left(-is\mu^2\right)}{(2\pi i s )}\exp\left(-\frac{\rho^2+\rho'^2}{4i s }\right) \int_{0}^{\infty}\frac{d\nu}{\pi} e^{-i\nu \tilde{\tau}}\sinh(\pi\nu)K_{i\nu}\left(\frac{\rho \rho'}{2i s }\right)\right]
	\end{align}
	where, $\mu^2=m^2+|\mathbf{k}_{\perp}|^2$. To simplify this expression further, we need the following  identity:
	\begin{align}
	\int_{0}^{\infty}\frac{du}{2s}\exp\left(-u-\frac{\rho_1^2+\rho_2^2}{4u}\right)K_{\nu}\left(\frac{\rho_1\rho_2}{2u}\right)&=K_{\nu}(\rho_1)K_{\nu}(\rho_2)
	\end{align}
	By a straightforward application of this result, we can simplify Eq.(\ref{GRexp}) to arrive at
	\begin{equation}\label{green_rind}
	G_{\mathcal{R}}(\mathbf{k}_{\perp},\tilde{\tau})=\int_{0}^{\infty}\frac{d\nu}{\pi^2}e^{-i\nu |\tau-\tau'|}\sinh(\pi\nu)K_{i\nu}(\mu\rho)K_{i\nu}(\mu\rho').
	\end{equation} 
	which matches with the known expressions in the literature (see for example, \cite{cr,linet}). A useful mnemonic is worth mentioning here; the $G_{\mathcal{R}}(\mathbf{k}_{\perp},\tilde{\tau})$ for a scalar field of mass $m$ in d+1 dimensions can be obtained from the propagator for a scalar field of mass $m$ in 1+1 dimension by the replacement: $m\rightarrow\mu$. 
 
  \section{Solving the Euclidean heat kernel equation}\label{heatkernel}

The line interval in the 2-dimensional plane can be expressed in polar and Cartesian coordinates in the form: 
\begin{equation}
ds^2 = dx^2 + dy^2 = r^2\ d\theta^2 + dr^2
\end{equation} 
As is well known, these are the Euclidean versions of the corresponding line element which we encounter in Minkowski and Rindler coordinates:
\begin{equation}
ds^2  = - dt^2 + dx^2  =- \rho^2d\tau^2+d\rho^2
\end{equation} 
The Euclidean quantum field theory can be reformulated in terms of an appropriate `heat kernel', the Euclidean analogue of Schwinger kernel, and is defined by: 
\begin{align}\label{Heat}
\frac{\partial K}{\partial s}-\Box_{E} K=\delta(s)\delta(x,x_0);\qquad \lim_{s\rightarrow 0}K=\delta(x,x_0)
\end{align}
Naively, one would have expected that the heat kernel obtained in the Euclidean sector of Rindler and Minkowski coordinates to match with the that in Cartesian and polar coordinates, respectively. 

The question then arises as to why the Rindler heat kernel $K^{Eu}_{\mathcal{R}}$ -- the Euclidean continuation of Schwinger kernel corresponding to the Rindler vacuum (and the corresponding Rindler propagator $G^{Eu}_{\mathcal{R}}$) -- is distinct from the Minkowski heat kernel $K^{Eu}_{\mathcal{M}}$ -- the Euclidean continuation of the Schwinger kernel corresponding to the Minkowski vacuum (and the Minkowski propagator $G^{Eu}_{\mathcal{M}}$) \cite{note6}. Kernels with non-trivial boundary conditions for the diffusion as well as the Schr\"{o}dinger equations in flat Euclidean space have been studied extensively in the literature ( for example, \cite{carslaw1910green,carslaw1920diffraction,Dowker:1977zj}). However, we review the solutions of heat kernel equation to trace how the distinction between different vacua are encoded in the boundary conditions. The formal solution to Eq.(\ref{Heat}), incorporating the correct boundary condition, can be given as
\begin{equation}
K = \exp(s\Box_{E}) [\delta(x,x_0)]
\end{equation} 
where $\delta(x,x_0)$ is a properly densitised Dirac delta function. Note that both the operator $\Box_{E}$ and $\delta(x,x_0)$ are generally covariant, making $K$ a covariant biscalar. If we solve this equation in one coordinate system (say, in the Cartesian coordinates) and transform everything covariantly to a new coordinate system ( say, polar coordinates), we will only get back the same heat kernel (and Feynman propagator) expressed in terms of the new coordinates. To get something like the (Euclidean) Rindler kernel, in this approach, we have to use two \textit{distinct} forms of Dirac delta functions in Cartesian and polar coordinate systems. We will now see in some detail what this implies. 

The basic difference between the approaches that lead to $K_{\mathcal{M}}$ and $K_{\mathcal{R}}$ can be understood through different representations of the Dirac delta functions. 
To see what is involved, recall that, using the properties of Dirac delta function, we can write
\begin{equation}
\delta(x-x')\delta(y-y')=\delta(r\cos\theta-r'\cos\theta')\delta(r\sin\theta-r'\sin\theta')
= \frac{\delta(r-r')}{r}\delta\left(2\sin\left[\frac{(\theta-\theta')}{2}\right]\right)
\label{deltatrans}
\end{equation} 
We now want to re-express this result as a Dirac delta function involving the $\theta$ coordinates. Using the basic property of Dirac delta function 
\begin{equation}
\delta(f(x))=\sum_{i}\frac{\delta(x-x_i)}{|f'(x_i)|}
\end{equation} 
and the periodicity of sine function, it immediately follows that 
\begin{equation}
\delta(x-x')\delta(y-y')=\sum_{n=-\infty}^{\infty}\frac{\delta(r-r')}{r}\delta(\theta-\theta'-2\pi n)
\label{k9}
\end{equation} 
In other words, the Dirac delta function in polar coordinates is an infinite sum. In fact, it is similar to what appears in our thermalisation of functions encountered in earlier sections. We could even say that when we proceed from the Cartesian coordinates to the polar coordinates the new Dirac delta function is obtained by thermalising $\delta (\theta -\theta')$. 

Since \eq{solk} gives the kernel as a result of a linear operation on the delta function, it immediately follows that each of the terms in \eq{k9} will lead to a kernel $K_n$ parameterized by the integer $n$.  The full kernel is provided by the infinite sum over these kernels $K_n$. To see how this comes about from first principles, let us represent the delta function in polar coordinates in \eq{k9} in terms of an appropriately chosen mode functions $\phi_m(\lambda;r,\theta)$ such that 
\begin{equation}
\sum_{m=-\infty}^{\infty}\int_{0}^{\infty}\phi_{m}(\lambda;r,\theta)\phi^*_{m}(\lambda;r',\theta')d\lambda=\sum_{n=-\infty}^{\infty}\frac{\delta(r-r')}{r}\delta(\theta-\theta'-2\pi n)
\label{k12}
\end{equation} 
where the functions 
\begin{equation}
\phi_{m}(\lambda;r,\theta)=\frac{e^{im\theta}}{\sqrt{2\pi}}\sqrt{\lambda}J_{|m|}(\lambda r)
\label{k13}
\end{equation} 
are eigenfunctions of the $\Box_{E}$ operator with eigen value $-\lambda^2$. 
If we now use \eq{solk} with this eigenfunction expansion, we will obtain 
\begin{equation}
\exp(s\Box_{E})\left[\sum_{n=-\infty}^{\infty}\frac{\delta(r-r')}{r}\delta(\theta-\theta'-2\pi n)\right]=\sum_{m=-\infty}^{\infty}\int_{0}^{\infty}e^{-\lambda^2s}\phi_{m}(\lambda;r,\theta)\phi^*_{m}(\lambda;r',\theta')d\lambda
\end{equation} 
Using the identities \cite[Eq.(10.22.67)]{DLMF} 
\begin{equation}
\int_{0}^{\infty}e^{-\lambda^2s}J_{m}(\lambda r)J_{m}(\lambda r')\lambda d\lambda=\frac{1}{2s}\exp\left(-\frac{r^2+r'^2}{4s}\right)I_{m}\left(\frac{rr'}{2s}\right)
\label{k17}
\end{equation} 
and \cite{watson1995treatise}
\begin{equation}
e^{a\cos z}=\sum_{m=-\infty}^{\infty}I_{m}(a)e^{im z}
\label{k19}
\end{equation} 
with $a=rr'/2s$ and $z=\theta - \theta'$, we get the final result
\begin{equation}
K_{\mathcal{M}}^{Eu}= \exp(s\Box)\left[\sum_{n=-\infty}^{\infty}\frac{\delta(r-r')}{r}\delta(\theta-\theta'-2\pi n)\right]=\frac{1}{(4\pi s)}\exp\left[-\frac{(r^2+r'^2-2rr'\cos(\theta-\theta'))}{4s}\right]
\end{equation} 
which, of course, matches with the result  obtained by transforming the standard Minkowski heat kernel $K_{\mathcal{M}}^{Eu}$ from the Cartesian coordinates to polar coordinates, treating it as biscalar.
So, by this procedure, we have just verified --- the rather trivial --- property of the kernel viz., it is generally covariant. 

There is another way of working with \eq{k9}. We can use the rule that we will confine to angular differences which satisfies $|\theta -\theta'| < 2\pi$. In that case only the $n=0$ term in \eq{k9} contributes and the expression for Dirac delta function becomes
\begin{equation}
\delta(x-x')\delta(y-y')=\frac{\delta(r-r')}{r}\delta(\theta-\theta')
\label{k9new}
\end{equation}
With this interpretation we will get back the same result \textit{as long as we keep `$m$' in the eigenfunction in \eq{k13} to be an integer.} In that case, the eigenfunctions are clearly periodic under $\theta \to \theta +2\pi$ and this is the key reason for us getting back $K^{Eu}_{\mathcal{M}}$. In this procedure, the mode functions will satisfy \eq{k12} with only the $n=0$ term being present in the right-hand side. 

To obtain the analogue of Rindler kernel we have to get out of the implicit periodicity in $\theta$.  This can be achieved as follows. Suppose we insist that the relevant Dirac delta function should not be taken as the one in \eq{k9} but with just the $n=0$ term. 
Further, we assume that the eigenvalue $m$ now takes all real values rather than integer values. That is, we will assume that we can replace the eigenfunction $\phi_m$  by $\phi_\omega$ with $\omega$ taking all real values. (This is, of course, essential for breaking the periodicity; if we take the eigenfunctions with $e^{im\theta}$ with integer values for $m$, they are formally periodic, even if we assume that $0\leq\theta<2\pi$.)
It would be interesting to ask how the previous analysis will change and what kind of kernel one will get. The entire analysis goes through with the new eigenfunction $\phi_\omega$. The completeness relation \eq{k12} now gets replaced by 
\begin{align}
\int_{-\infty}^{\infty}d\omega\int_{0}^{\infty}d\lambda\phi_{\omega}(\lambda;r,\theta)\phi^{*}_{\omega}(\lambda;r',\theta')&=\frac{\delta(r-r')}{r}\    \int_{-\infty}^{\infty}\frac{d\omega}{2\pi} e^{i\omega(\theta-\theta')}\\
&=\frac{\delta(r-r')}{r}\delta{(\theta-\theta')}
\end{align}
This will lead to the Rindler kernel, $K^{Eu}_{\mathcal{R}}$: 
\begin{align}
K^{Eu}_{\mathcal{R}}&=\exp(s\Box_{E})\frac{\delta(r-r')}{r}\delta{(\theta-\theta')}\\
&=\int_{-\infty}^{\infty}d\omega\int_{0}^{\infty}d\lambda\, e^{-\lambda s^2}\phi_{\omega}(\lambda;r,\theta)\phi^{*}_{\omega}(\lambda;r,\theta)
\end{align}
which, using the definition of $\phi_\omega$ from \eq{k13} and the identity in \eq{k17}, can be expressed as 
\begin{equation}\label{rind_kernel}
K^{Eu}_{\mathcal{R}}=\frac{1}{(4\pi s)}\exp\left(-\frac{r^2+r'^2}{4s}\right)\int_{-\infty}^{\infty}d\omega I_{|\omega|}\left(\frac{rr'}{2s}\right)e^{i\omega(\theta-\theta')}
\end{equation} 
From the known asymptotic behaviour of $I_\nu(z)$ one can directly verify that this kernel indeed satisfies the boundary condition we started with, namely:
\begin{equation}
\lim_{s\rightarrow 0}K^{Eu}_{\mathcal{R}}=\frac{\delta(r-r')}{r}\delta(\theta-\theta')
\end{equation} 

In other words, the expression for the kernel in \eq{solk} correctly reproduces both $K^{Eu}_{\mathcal{R}}$ and $K^{Eu}_{\mathcal{M}}$ depending on how we interpret the Dirac delta function on the right-hand side. If we take the ``natural'' option and insist that $\theta$ is an angle with periodicity $2\pi$ then we will \textit{only} get $K^{Eu}_{\mathcal{M}}$, which, of course, has this periodicity arising from the thermalizing condition \eq{thermaltpK}. In order to get $K^{Eu}_{\mathcal{R}}$ one has to \textit{explicitly break} this periodicity condition in the Euclidean sector (also see note\cite{note7}). 
It, therefore, appears that in the Euclidean sector $K^{Eu}_{\mathcal{M}}$ arises rather naturally while we need to do something artificial to get $K^{Eu}_{\mathcal{R}}$, namely allowing the kernel to be multivalued in the Euclidean plane. However, $K^{Eu}_{\mathcal{R}}$ does arise naturally as the heat kernel in an infinitely sheeted, locally flat 2-d Riemann surface. 

Incidentally, the situation is somewhat different in the Lorentzian sector. If we do the coordinate transformation of the  Dirac delta function in the Lorentzian sector, then instead of 
\eq{deltatrans} we will get an expression with 
$\delta [ 2\sinh \{(\tau - \tau')/2\}]$. This leads to just $\delta [\tau - \tau']$ as long as we stick to real values of $\tau$ and $\tau'$, as we should.  (The periodicity now is in the imaginary values of $\tau-\tau'$ which do not play a role as long as we decide to live in Lorentzian spacetime.) This would suggest that an analysis similar to what we did above will not recognize any periodicity in the Lorentzian sector. As a result, this will lead to the Rindler kernel $K_{\mathcal{R}}$ in a natural fashion. To obtain the Minkowski kernel $K_{\mathcal{M}}$ in the Lorentzian sector, working entirely in Rindler coordinates, we have to thermalize $K_{\mathcal{R}}$ explicitly. We find it rather intriguing that the natural solution to \eq{solk} leads to the Rindler kernel $K_{\mathcal{R}}$ in the Lorentzian sector while it leads to the Minkowski kernel $K_{\mathcal{M}}$ in the Euclidean sector. We will come back to this feature in the last section. 

It can be directly verified that the Euclidean kernels $K^{Eu}_{\mathcal{R}}$ and $K^{Eu}_{\mathcal{M}}$, upon integration over $s$, leads to the correct propagators $G^{Eu}_{\mathcal{R}}$ and $G^{Eu}_{\mathcal{M}}$, respectively. This can be obtained most easily by first rewriting the relevant expressions in terms of $K_{i\nu}$ in the following forms. 
\begin{equation}
\label{EuclideanKM}
K^{Eu}_{\mathcal{M}}=\frac{1}{(2\pi s )}\exp\left(-\frac{r^2+r'^2}{4 s }\right)\int_{0}^{\infty}\frac{d\nu}{\pi}\cosh[\nu(\pi-\phi)]K_{i\nu}\left(\frac{rr'}{2s}\right)
\end{equation} 
and 
\begin{equation}
K^{Eu}_{\mathcal{R}}=\frac{1}{(2\pi s)}\exp\left(-\frac{r^2+r'^2}{4s}\right) \int_{0}^{\infty}\frac{d\nu}{\pi} e^{-\nu \phi}\sinh(\pi\nu)K_{i\nu}\left(\frac{rr'}{2s}\right)
\end{equation} 
These expressions are derived in Appendix \ref{AppendixC}. The integration leads to the following expressions in the Euclidean sector, which are well known in the literature:
\begin{equation}
G^{Eu}_{\mathcal{R}}=\int_{0}^{\infty}\frac{d\nu}{\pi^2}e^{-\nu |\theta-\theta'|}\sinh(\pi\nu)K_{i\nu}(mr)K_{i\nu}(mr')
\end{equation}
and 
\begin{align}\label{gmeu}
G^{Eu}_{\mathcal{M}}&=\int_{0}^{\infty}\frac{d\nu}{\pi^2}\cosh[\nu(\pi-|\theta-\theta'|)]K_{i\nu}(mr)K_{i\nu}(mr')
\end{align}
This matches exactly with the previous results, for example, that in \cite{linet}.

\section{From the Euclidean Plane to the four Rindler wedges}

Finally, we will take up the third motivation for this work, mentioned in Section \ref{sec:mot3}. We will show that the Lorentzian biscalars (like $K_{\mathcal{M}}(x_1,x_2;s)$ and $G_{\mathcal{M}}(x_1,x_2)$) can indeed be obtained from their corresponding Euclidean avatars ( $K^{Eu}_{\mathcal{M}}(x_1,x_2;s)$ and $G^{Eu}_{\mathcal{M}}(x_1,x_2)$) by analytic continuation. However the standard analytic continuation, discussed in the literature --- which involves replacing $t_E\to -it$ and $\tau_E\to -i\tau$ in $x=\rho\cos\tau_E, t_E=\rho\sin\tau_E$ --- will only lead us to the events in the right Rindler wedge. To obtain the form of the biscalars in other wedges, we actually need different sets of analytic continuations. This is particularly important when the two events in the biscalars are located in different wedges. 

The key (unifying) principle which helps us to discover the correct analytic continuation is the following: The invariant Euclidean distance squared ($\sigma_{E}^2$) should be analytically continued to the invariant Lorentzian distance squared ($\sigma^2$) such that the latter has an infinitesimal imaginary part. Depending on the location of the two events that define the Lorentzian kernel, different cases lead to different mappings which we will list below. We first discuss the analytic continuation of the kernel and then describe the corresponding results for the propagator.

\subsection{The recipe for analytic continuation}\label{KM_Eu_to_Mink}
The analytic continuation from Euclidean to Minkowski space for simple scalar functions was discussed in \cite{Padmanabhan:2019yyg}. We generalize this prescription and propose a recipe for analytically continuing kernels and Feynman propagators, which are biscalars. To accomplish this, it suffices to analytically continue two basic biscalars in Euclidean space: (i) $\sigma^2_{E}$, the square of Euclidean distance and $\Theta_{E}=|\theta-\theta'|$, the angle between vectors, say $\mathbf{x}$ and $\mathbf{x}'$, representing the two points. To reproduce the analytical structure of Feynman propagator and kernel, we should ensure that $\sigma^2_{E}$ be analytically continued to $\sigma^2+i 0^{+}$ in the Lorentzian sector. To analytically continue $\Theta_{E}$, let us start by writing its explicit form:
\begin{align}
\Theta_{E}(\mathbf{x},\mathbf{x}')&=\cos^{-1}\left[\frac{\mathbf{x}\cdot\mathbf{x}'}{|\mathbf{x}||\mathbf{x}'|}\right]
\end{align}  
where, $\mathbf{x}$ and $\mathbf{x}'$ are the Cartesian coordinates of two points in the Euclidean plane and dot denotes the standard inner product in the Euclidean flat space. It is convenient at this stage to define the quantity in the square bracket as $Z(\mathbf{x},\mathbf{x}')$. In terms of the polar coordinates, the biscalar $Z$ can be written as
\begin{align}
Z(\mathbf{x},\mathbf{x}')=\frac{r^2-r'^2-\sigma_{E}^2}{2rr'}.
\end{align} 
Note that $\Theta_{E}$ is a function of $Z$ alone; more specifically we have, $\Theta_{E}=\cos^{-1}(Z)$. However, since we are interested in analytic continuation of $\Theta_{E}$, we need an expression for $\Theta_E(Z)$ that is valid in the whole of complex $Z-$plane. For this purpose, recall that the expression for $\cos^{-1}(y)$, for real $y$ in the range $[-1,1]$ is given by
\begin{align}
\cos^{-1}(y)&=\frac{\pi}{2}-\sin^{-1}(y)
=\frac{\pi}{2}+i\log(iy+\sqrt{1-y^2})
\end{align}
which is easy to verify (see for example \cite{note8}). This expression enables us to analytically continue the $\cos^{-1}$ function into the complex plane. However, one cannot naively extend it into an analytic function in the entire complex place, since it inherits branch cuts from the `$\log$' and the `square root function' that appear in its definition. Hence, a complex extension of $\Theta_{Z}$ needs to be defined using appropriate $i\epsilon$ prescription as:   
\begin{align}\label{defTheta}
\Theta_{E}(Z)=\frac{\pi }{2}+i \log\left(\sqrt{1-Z^2+i0^{+}}+i Z\right)
\end{align}
With these two inputs we get a consistent analytic continuation of the Euclidean biscalars $K^{Eu}_{\mathcal{M}}$ and $G^{Eu}_{\mathcal{M}}$ to the whole of Minkowski space. We give the explicit recipe in Table \ref{tab:table1} for three cases: (i) $RR$, when both points are in $R$ wedge (ii) $FF$, when both points are in $F$ wedge and (iii) $RF$, when one point is in $F$ and the other in $R$. (More algebraic details  are given in Appendix \ref{appendixd}). The other cases (involving L and P) can be found in a similar manner.

\begin{table}[h!]
	\begin{center}
		\caption{Recipe for analytic continuation}
		\label{tab:table1}
		\begin{tabular}{|c|c|c|c|}
			\hline
			\textbf{Case} & $\textrm{Euclidean}\rightarrow \textrm{Lorentzian}$ & \textbf{$\sigma_{E}^2\rightarrow\sigma^2$} & \textbf{$\Theta_{E}\rightarrow\Theta$}\\ 
			\hline
			$RR$ & $(r,\theta)\rightarrow(\rho,i\tau e^{-i \epsilon})$
			& $\rho^2+\rho'^2-2\rho\rho'\cosh(\tau-\tau')+i0^{+}$ & $i|\tau-\tau'|+0^{+}$\\ 
			~~&$(r'\theta')\rightarrow(\rho',i\tau'e^{-i\epsilon})$&~~&~~\\
			\hline% <--
			$RF$ & $(r,\theta)\rightarrow(\rho_R,i\tau_{R})$ &$-\rho_{F}^2+\rho_{R}^2-2\rho_{F}\rho_{R}\sinh(\tau_{F}-\tau_{R})+i0^{+}$ & $i(\tau_{F}-\tau_{R})+\frac{\pi}{2}+\epsilon$ \\
			~~&$(r',\theta')\rightarrow(i\rho_{F},i\tau_{F}+\frac{\pi}{2}+\epsilon)$&~~&~~\\
			\hline % <--
			$FF$ & $(r_{<},\theta)\rightarrow(-e^{i\epsilon}i\rho_{<},i\tau+\frac{\pi}{2})$ & $-\rho_{<}^2-\rho_{>}^2+2\rho_{<}\rho_{>}\cosh(\tau-\tau')+i0^{+}$ & $-i|\tau-\tau'|+\pi$\\
			~~&$(r_{>},\theta')\rightarrow (i\rho_{>},i\tau'-\frac{\pi}{2})$&~~&~~ \\
			\hline% <--
		\end{tabular}
	\end{center}
\end{table}

\subsection{Analytic continuation of $K_{\mathcal{M}}^{Eu}$ and $G_{\mathcal{M}}^{Eu}$}
With the recipe in Table. \ref{tab:table1} and the analytic continuation of the Schwinger propertime  $s \rightarrow ise^{-i0^{+}}$, we can  obtain the expressions for $K_{\mathcal{M}}$ and $G_{\mathcal{M}}$,  by analytic continuation of $K_{\mathcal{M}}^{Eu}$ and $G_{\mathcal{M}}^{Eu}$. We will start with $K_{\mathcal{M}}$ in the $RR$, $RF$ and $FF$ sectors. (We use the notation $AB$ to describe the situation in which the two events are in the wedges $A$ and $B$, respectively.) The Feynman propagator can then be found using the standard relation
\begin{align}\label{gfromk}
i G_{\mathcal{M}}(x,x')=\int_{0}^{\infty}ds\ e^{-im^2s}K_{\mathcal{M}}(x,x';s)
\end{align}
It is clear that we will encounter the following type of integral, while evaluating $G_{\mathcal{M}}$ using the above relation (after rotating the contour of integration via $s\rightarrow-is$):
\begin{align}
\mathcal{I}(z_1,z_2)=\int_{0}^{\infty}\frac{ds}{2s} e^{-m^2s} \exp\left(-\frac{z_1^2+z_2^2}{4s}\right)K_{i\nu}\left(\frac{z_1z_2}{2s}\right)
\end{align} 
When $z_1$ and $z_2$ satisfies the following conditions,
\begin{align}
|\arg[z_1]|<\pi;\qquad|\arg[z_2]<\pi|;\qquad|\arg[z_1+z_2]|<\frac{\pi}{4},
\end{align}
one can evaluate the integral explicitly to get,
\begin{align}\label{integralresult}
\mathcal{I}(z_1,z_2)=K_{i\nu}(mz_1)K_{i\nu}(mz_2).
\end{align}
Our strategy will be to analytically continue $\mathcal{I}(z_1,z_2)$ for the desired values of $z_1$ and $z_2$ by using the appropriate analytic continuation of $K_{i\nu}$. We will now work out the calculations explicitly for the RR, FF and RF cases. (These results have been obtained earlier in Ref.\cite{Boulware:1974dm} by a more complicated procedure.)

\subsubsection{RR: both events in the $R$ wedge}
Application of our prescription to Eq.(\ref{EuclideanKM}) yields the following expression for Minkowski kernel $K^{(m)}_{\mathcal{M}}$ for a massive scalar field, when both events are in the $R$ wedge:
\begin{align}\label{kmrr}
K_{\mathcal{M}}^{(m)}&=\frac{\exp\left(-i\mu^2s\right)}{(4\pi is )}\exp\left(-\frac{\rho^2+\rho'^2}{4 is} \right)\int_{-\infty}^{\infty}\frac{d\nu}{\pi} e^{- i\nu\left[|\tau-\tau'|+i (\pi-\epsilon)\right]}K_{i\nu}\left[\frac{\rho\rho'}{2is}\right]
\end{align}
This is for a $(1+d)$ dimensional, massive case and --- as discussed earlier in connection with \eq{k86gen} --- we have taken a Fourier transform with respect to the transverse coordinate differences and set $\mu^2=m^2+k_\perp^2$.
(A similar result can be obtained when both points are in the $L$-wedge, which we will not discuss explicitly.) The Feynman propagator can now be found by using \eq{gfromk} and \eqref{integralresult} to get:
\begin{align}\label{gmrr}
G_{\mathcal{M}}=\frac{1}{2\pi^2}\int_{-\infty}^{\infty}e^{- i\nu\left[|\tau-\tau'|+i (\pi-\epsilon)\right]}K_{i\nu}(\mu\rho)K_{i\nu}(\mu\rho')\,d\nu
\end{align}
To arrive at a more familiar form of $K_{\mathcal{M}}$, we can use the following result:
\begin{align}\label{relationk0}
\int_{-\infty}^{\infty}e^{-i(\xi_{R}+i\xi_{I})\omega}K_{i\omega}(\alpha)K_{i\omega}(\beta)\,d\omega=\pi K_{0}\left[\sqrt{\alpha^2+\beta^2+2\alpha\beta\cosh(\xi_{R}+i\xi_{I})}\right];
\end{align}
which is valid for 
\begin{equation}
|\arg[\alpha]|+|\arg[\beta]|+|\xi_{I}|<\pi 
\label{validity}
\end{equation} 
With $\alpha=\mu\rho$, $\beta=\mu\rho'$, $\xi_{R}=|\tau-\tau'|$ and $\xi_{I}=(\pi-\epsilon)$, we arrive at:
\begin{align}\label{gmrr2}
G_{\mathcal{M}}=\frac{1}{2\pi}K_{0}\left(\mu\sqrt{\rho^2+\rho'^2-2\rho\rho'\cosh(\tau-\tau')+i0^{+}}\right)
\end{align}
which gives the correct propagator for a massive scalar field (see e.g.\cite{linet}). The following point needs to be emphasized in the above calculation: We need a negative sign for the $\cosh$ term in the argument of $K_0$ to give the correct propagator while \eq{relationk0} has a positive sign for the $\cosh$ term in the argument of $K_0$. We can flip this sign if we could take $\xi_I=\pi$, but this is forbidden due to the condition of validity, given by \eq{validity}. We can, of course,  set $\xi_{I}=(\pi-\epsilon)$ to flip the sign of the $\cosh$ term in the argument of $K_0$, 
which is what we have done.
However, this procedure adds an infinitesimal imaginary part --- coming from $\epsilon$--- to the argument, which is indicated by $i0^+$ in \eq{gmrr2}. Nevertheless, this is exactly what we need in the Feynman propagator, and thus everything works out consistently, though somewhat subtly.

There is another representation of this propagator which is useful. This representations is motivated by the fact that \eq{gmrr} contains $|\tau-\tau'|$ while it will be useful to have just the Fourier transform with $(\tau-\tau')$. (The Fourier transform of the propagator with respect to $(\tau-\tau)$ can be interpreted as the amplitude for propagation in energy space and is closely related to the thermality of Rindler horizon \cite{Padmanabhan:2019yyg}). This can be achieved as follows:
Using the result Eq.(\ref{relationk0}) again, but this time with $\xi_{R}=(\tau-\tau'),\xi_{I}=0,\alpha=e^{i(\pi-\epsilon)}\mu\rho_{<}$ and $\beta=\mu\rho_{>}$ we arrive at a different representation for $G_{\mathcal{M}}$, given by
\begin{align}\label{gmrr3}
G_{\mathcal{M}}=\frac{1}{2\pi^2}\int_{-\infty}^{\infty}e^{- i\nu(\tau-\tau')}K_{i\nu}(e^{i(\pi-\epsilon)}\mu\rho_{<})K_{i\nu}(\mu\rho_{>})d\nu
\end{align}
This expression reveals that, the Fourier transform $\tilde{G}_{\mathcal{M}}(\nu)$ (with the dependence on other variables being suppressed) of $G_{\mathcal{M}}$ with respect to $(\tau-\tau')$ is given by
\begin{align}\label{ftgm}
\tilde{G}_{\mathcal{M}}(\nu)=\frac{1}{\pi}K_{i\nu}(e^{i(\pi-\epsilon)}\mu\rho_{<})K_{i\nu}(\mu\rho_{>})
=\frac{1}{\pi}K_{i\nu}(-\mu\rho_{<})K_{i\nu}(\mu\rho_{>})
\end{align}
where the $i\epsilon$ is suppressed in the last equality.
This result was derived in \cite{Padmanabhan:2019yyg} by a different procedure. 

It is also possible to obtain a similar Fourier transform for the `reflected' propagator, that will be useful later on. This uses another choice of parameters in Eq.(\ref{relationk0}) corresponding  to $\alpha=\mu\rho,\beta=\nu\rho',\xi_{R}=(\tau-\tau')$ and $\xi_{I}=0$. A direct substitution of this choice into the RHS of Eq.(\ref{relationk0}) gives $\pi K_{0}(\sqrt{\rho^2+\rho'^2+2\rho\rho'\cosh(\tau-\tau')})$. A comparison of this expression (divides by $2\pi^2$) with Eq.(\ref{gmrr2}) shows that it corresponds to $G_{\mathcal{M}}^{(r)}$, the Feynman propagator between $(\rho,\tau)$ and another event in $L$ obtained by reflecting $\rho',\tau'$ about the origin of $x-t$ plane. Hence, again using Eq.(\ref{relationk0}), with our latest choice of parameters, gives us
\begin{align}\label{ftgmr}
\tilde{G}_{\mathcal{M}}^{(r)}(\nu)&=\frac{1}{\pi}K_{i\nu}(\mu\rho)K_{i\nu}(\mu\rho')
\end{align}
This result, along with Eq.(\ref{gmrr3}), will be made use later for a simple derivation of the Rindler propagator $G_{\mathcal{R}}$. 

Another important techinal remark concerning the structure of $\tilde{G}_{\mathcal{M}}(\nu)$ and $\tilde{G}^{(r)}_{\mathcal{M}}(\nu)$ in \eq{ftgm} and \eq{ftgmr} is in order. We know that, the propagator $G_{\mathcal{M}}$ and the `reflected' propagator $G_{\mathcal{M}}^{(r)}$ are related by $G_{\mathcal{M}}^{(r)}(\tau-\tau')=G_{\mathcal{M}}(\tau-\tau'+i\pi)$ involving the shift of the argument by an imaginary quantity. A naive (wrong) application of the shift theorem for the Fourier transform (valid for shifts by real quantities) will now give, $\tilde{G}^{(r)}_{\mathcal{M}}(\nu)=e^{\pi\nu}\tilde{G}_{\mathcal{M}}(\nu)$ which, as we can see from \eq{ftgm} and \eq{ftgmr}, is incorrect. As we have mentioned  earlier --- see the discussion after \eq{ftrelation} ---
the Fourier transforms of two functions $f(x)$ and $f_{y}(x)\equiv f(x+iy)$, with respect to $x$, are in general \textit{not} related by $\tilde{f}_{y}(\nu)=e^{y\nu}\tilde{f}(\nu)$, for real $y$. The Fourier transform of $f(x)$ exists and is give by $\tilde{f}(\nu)$, if and only if the following integral converges:
\begin{align}
\int_{-\infty}^{\infty}\tilde{f}(\nu) e^{-i\nu x}\frac{d\nu}{2\pi} =f(x)
\label{conv1}
\end{align}
However, the convergence of this integral in \eq{conv1} does not guarantee that the following integral also converges, for an arbitrary real $y$:
\begin{align}
I=\int_{-\infty}^{\infty}\left[e^{y\nu}\tilde{f}(\nu)\right] e^{-i\nu x}\frac{d\nu}{2\pi}
\label{conv2}
\end{align}
For the  integral in \eq{conv2} to converge, $\tilde{f}(\nu)$ should satisfy the additional condition that $e^{y\nu}\tilde{f}(\nu)$ vanishes sufficiently fast as $\nu\rightarrow \pm\infty$, where the $+/-$ sign is for positive/negative values of $y$, respectively. Hence, in general, $e^{y\nu}\tilde{f}(\nu)$ will not be the Fourier transform of $f_{y}(x)$ and one cannot obtain the result by shifting.

Let us illustate this aspect in the  case of the Minkowski propagators $G_{\mathcal{M}}$ and $G^{(r)}_{\mathcal{M}}$. Recall that we derived the Fourier transform of $G_{\mathcal{M}}$ using the relation \eq{relationk0}, with the parameters being chosen as $\xi_{R}=(\tau-\tau'),\xi_{I}=0,\alpha=e^{i(\pi-\epsilon)}\mu\rho_{<}$ and $\beta=\mu\rho_{>}$. Since, $G_{\mathcal{M}}(\tau-\tau+i\pi)=G_{\mathcal{M}}^{(r)}$, one may naively chose $\xi_{R}=(\tau-\tau'),\xi_{I}=i\pi,\alpha=e^{i(\pi-\epsilon)}\mu\rho_{<}$ and $\beta=\mu\rho_{>}$ to `derive' the Fourier transform $\tilde{G}_{\mathcal{M}}^{(r)}$. Using this choice in \eq{relationk0} would then lead to the  wrong result, given by shifting:
$
\tilde{G}_{\mathcal{M}}^{(r)}(\nu)=e^{\pi\nu}\tilde{G}_{\mathcal{M}}(\nu).
$
This result is wrong because, for this choice of parameters,  the left hand side \eq{validity} is given by:
\begin{align}
|\arg[\alpha]|+|\arg[\beta]|+|\xi_{I}|=2\pi-\epsilon>\pi
\end{align}
thereby violating the condition in \eq{validity}. Since \eq{validity} is a necessary condition for using the integral in \eq{relationk0}, violating it leads to a wrong result. More precisely, our choice of parameters 
does not satisfy the necessary condition for the integral in \eq{relationk0} to converge, thereby invalidating the procedure of shifting the contour in the complex plane; this is the reason why  $\tilde{G}_{\mathcal{M}}^{(r)}(\nu)\neq e^{\pi\nu}\tilde{G}_{\mathcal{M}}(\nu)$. To derive the correct Fourier transform of $G_{\mathcal{M}}^{(r)}$, one has to proceed exactly as we did earlier; by choosing 
$\alpha=\mu\rho,\beta=\nu\rho',\xi_{R}=(\tau-\tau')$ and $\xi_{I}=0$. With this choice of parameters, the right hand side of \eq{relationk0} takes the required form that is relevant for $G_{\mathcal{M}}^{(r)}$ and the condition $|\arg[\alpha]|+|\arg[\beta]|+|\xi_{I}|<\pi$ is also satisfied (see \cite{note9} for a different context in which this condition is considered). This leads to a convergent integral and the correct Fourier transform in \eq{ftgmr}.

\subsubsection{FF: both events in the F wedge}
Once again, using Table \ref{tab:table1} in Eq.(\ref{EuclideanKM}) we get
\begin{align}
K_{\mathcal{M}}^{(m)}&=\frac{\exp\left(-i\mu^2s\right)}{(4\pi is )}\exp\left(\frac{\rho_{>}^2+\rho_{<}^2}{4 is }\right)\int_{-\infty}^{\infty}\frac{d\nu}{\pi} e^{- i\nu(\tau-\tau')}K_{i\nu}\left[\frac{e^{i\epsilon}\rho_{<}\rho_{>}}{2is}\right]
\end{align}    
In this case, we have replaced $|\tau-\tau'|$ by $(\tau-\tau')$ to arrive at this expression in the form of a Fourier transform. We could do this in this case --- but not in the case of \eq{kmrr} --- because the rest of the integrand is symmetric under $\nu\rightarrow-\nu$. The $\nu$ integration can be easily done to  reduce this expression to the familiar form:
\begin{align}
K_{\mathcal{M}}^{(m)}&=\frac{\exp\left(-i\mu^2s\right)}{(4\pi is)}\exp\left[-\frac{1}{4 is }(-\rho^2-\rho'^2+2\rho\rho'\cosh(\tau-\tau')+i0^{+})\right]
\end{align}
To find the expression for $G_{\mathcal{M}}$, we need to apply the following substitution in Eq.(\ref{integralresult}),
\begin{align}
z_1=e^{i\left(-\frac{\pi}{2}+\epsilon\right)}\rho_{<};\qquad z_2=e^{i\frac{\pi}{2}}\rho_{>},
\end{align}
Further, using the standard connection formulas between the MacDonald and Hankel functions, we arrive at
\begin{align}\label{gminff}
G_{\mathcal{M}}=\frac{1}{4}\int_{-\infty}^{\infty}e^{-i\nu(\tau-\tau')}H^{(1)}_{i\nu}(\mu e^{i\epsilon}\rho_{<})H^{(2)}_{i\nu}(\mu\rho_{>})d\nu
\end{align}
This expression has a simple interpretation. 
It is well known that the  solutions to the massive Klein-Gordon equation in F, which  is a positive frequency mode with respect to the \textit{inertial time}  is given by (see for example \cite{Higuchi:2018tuk}):
\begin{align}
\psi_{\nu,\mathbf{k}_{\perp}}(\rho,\tau)\propto e^{\pi|\nu|/2} H^{(2)}_{i|\nu|}(\mu\rho)e^{-i\nu\tau}
\end{align}
In terms of these modes, with appropriate normalization, Eq.(\ref{gminff}) reduces to the suggestive form
\begin{align}\label{xxxx}
G_{\mathcal{M}}=\int_{-\infty}^{\infty}\frac{d\nu}{(2\pi)}\psi_{\nu,\mathbf{k}_{\perp}}(\rho_{>},\tau)\psi_{\nu,\mathbf{k}_{\perp}}^*(\rho_{<},\tau').
\end{align}
In the $F$ wedge,  $\rho$ acts as the time coordinate and $\tau$ acts as the space coordinate. So the subscripts $<$ and $>$  in the above expression actually give us a \textit{time-ordered} expression. 
So \eq{xxxx} has the correct form of the time ordered correlator in the Minkowski vacuum, expressed in terms of positive frequency mode functions, as is appropriate for the Feynman propagator.
\subsubsection{RF: one event in R and the other in F}
In this case, $K_{\mathcal{M}}$ takes the form:
\begin{align}
K_{\mathcal{M}}^{(m)}&=\frac{\exp\left(-i\mu^2s\right)}{(4\pi is )}\exp\left(-\frac{\rho_{R}^2-\rho_{F}^2}{4 is }\right)\int_{-\infty}^{\infty}d\nu\, e^{- i\nu(\tau_F-\tau_R)}\left\{\frac{e^{\frac{\pi\nu}{2}}}{\pi}K_{i\nu}\left[\frac{\rho_{F}\rho_{R}}{2s}\right]\right\}
\end{align}    
The term in curly brackets is just the Fourier transform of $e^{iz\sinh(\tau_F-\tau_R)}$ with respect to $\tau_F$, where $z=(\rho_{F}\rho_{R})/(2s)$. Therefore this expression may be simplified to get;
\begin{align}
K_{\mathcal{M}}^{(m)}&=\frac{\exp\left(-i\mu^2s\right)}{(4\pi i s)}\exp\left[-\frac{1}{4 is }(\rho_{R}^2-\rho_{F}^2+2\rho_F\rho_R\sinh(\tau_F-\tau_R)+i0^+)\right]
\end{align} 
Once again, we obtain the familiar expression. To find the Feynman propagator, in Eq.(\ref{integralresult}) choose $z_1$ and $z_2$ to be the following:
\begin{align}
z_1=\rho_R;\qquad z_2=e^{i\frac{\pi}{2}}\rho_{F}.
\end{align}
The form of $G_{\mathcal{R}}$ in this case can then be found to be
\begin{align}
G_{\mathcal{M}}=\frac{-i}{4\pi}\int_{-\infty}^{\infty}e^{-i\nu(\tau_F-\tau_R)}H_{i\nu}^{(2)}(\mu\rho_{F})K_{i\nu}(\mu\rho_R)\,d\nu.
\end{align}
In a recent work \cite{Padmanabhan:2019yyg}, this form of the Minkowski propagator was a key ingredient in a simple derivation of thermality of the Rindler horizon. 

\subsection{Analytic continuation of $K^{Eu}_{\mathcal{R}}$ and $G^{Eu}_{\mathcal{R}}$}
In exactly the same manner as we analytically continued the Euclidean-\textit{Minkowski} kernel and propagator, we can also analytically continue the Euclidean-\textit{Rindler} kernel and propagator. Purely algebraically, we can do this in all the four wedges. However, the Rindler vacuum state --- used implicitly in the construction of the Rindler  kernel and Rindler propagator --- has a natural definition only in $R$ and $L$ wedges. So we shall confine our analysis in this paper to the  two cases in which the analytic continuation of $K^{Eu}_{\mathcal{R}}$ and $G^{Eu}_{\mathcal{R}}$ has a natural interpretation viz., in $RR$ and $LL$. We will only present the $RR$ case here, since the $LL$ case can be dealt in a identical  manner. In $RR$ our procedure leads to the kernel:
\begin{align}
K_{\mathcal{R}}^{(m)}&=\frac{\exp\left(-i\mu^2s\right)}{(2\pi^2 is )}\exp\left(-\frac{\rho^2+\rho'^2}{4 is} \right)\int_{0}^{\infty}d\nu\, e^{- i\nu|\tau-\tau'|-\epsilon\nu}\left( \sinh \pi\nu\right)K_{i\nu}\left[\frac{\rho\rho'}{2is}\right]
\end{align} 
The Feynman propagator $G_{\mathcal{R}}$ can now be computed to give expressions previously known in literature (see e.g.\cite{cr,linet}):
\begin{align}\label{grexpr}
G_{\mathcal{R}}&=\frac{1}{\pi^2}\int_{0}^{\infty}e^{-i\nu|\tau-\tau'|}\sinh(\pi\nu)K_{i\nu}(\mu\rho)K_{i\nu}(\mu\rho')\,d\nu
\end{align}
The Rindler modes $u_{\nu,\mathbf{k}_{\perp}}(\rho,\tau)$, which are positive frequency solutions with respect to the Rindler time coordinate, are given by \cite{Crispino:2007eb}
\begin{align}
u_{\nu,\mathbf{k}_{\perp}}(\rho,\tau)=\frac{\sqrt{\sinh(\pi\nu)}}{\pi}K_{i\nu}(m\rho)e^{-i\nu\tau};\qquad\nu>0
\end{align}   
The Feynman propagator can then be rewritten in the expected form;
\begin{align}\label{grrr1}
G_{\mathcal{R}}&=\int_{0}^{\infty}d\nu\, u_{\nu,\mathbf{k}_{\perp}}(\rho,\tau_{>})u_{\nu,\mathbf{k}_{\perp}}(\rho',\tau_{<})
\end{align}   
where, $\tau_{>}>\tau_{<}$. This representation confirms that the propagator $G_{\mathcal{R}}$ that we derived is indeed a time ordered correlation function. 

Another important quantity that we are interested in is the Fourier transform of $G_{\mathcal{R}}$ with respect to $(\tau-\tau')$. We will first derive it from \eq{grexpr} and then show that our result Eq.(\ref{ftrelation}) reproduces it in an alternative, simple, manner. For the former derivation, let us first rewrite the integral in Eq.(\ref{grexpr}) as:
\begin{align}\label{gr2form}
G_{\mathcal{R}}&=\frac{1}{2\pi i}\int_{0}^{\infty}d\nu\, e^{-i\nu|\tau-\tau'|}I_{-i\nu}(\mu\rho_{<})K_{i\nu}(\mu\rho_{>})-\frac{1}{2\pi i}\int_{0}^{\infty} d\nu\, e^{-i\nu|\tau-\tau'|}I_{i\nu}(\mu\rho_{<})K_{i\nu}(\mu\rho_{>})
\end{align}
where, we have used the identity $(2i/\pi)\sinh(\pi\nu)K_{i\nu}(x)=I_{-i\nu}(x)-I_{i\nu}(x)$. Since, there are no poles for $I_{i\nu}(m\rho_{<})K_{i\nu}(m\rho_{>})$ in the lower half complex $\nu-$plane and the condition $\rho_{>}>\rho_{<}$ ensures that this terms vanishes sufficiently fast \cite{note1}  as $\textrm{Im}[\nu]\rightarrow-\infty$, we can rotate the $\nu$ integration in the second term of \eq{gr2form} to the straight line contour from $\nu=0$ to $\nu\sim-\infty e^{-i 0^{+}}$. Therefore, the expression for $G_{\mathcal{R}}$ reduces to the form
\begin{align}
G_{\mathcal{R}}=\frac{1}{2\pi i}\int_{-\infty}^{\infty}d\nu\, e^{-i\nu|\tau-\tau'|}I_{-i|\nu|}(\mu\rho_{<})K_{i\nu}(\mu\rho_{>})
\end{align}
Since the integrand in the last expression is symmetric under $\nu\rightarrow-\nu$, we can as well replace $|\tau-\tau|$ with $(\tau-\tau')$ to reveal the Fourier transform of $G_{\mathcal{R}}$ with respect to $(\tau-\tau')$. Hence, we obtain
\begin{align}
\tilde{G}_{\mathcal{R}}=-iI_{-i|\nu|}(\mu\rho_{<})K_{i\nu}(\mu\rho_{>})
\end{align} 
which is the Fourier transform of the Rindler propagator. As far as we know this simple expression has not been obtained before in the literature.

We shall now present an alternative, simpler,  derivation of this result by making use of Eq.(\ref{ftrelation}). Note that, since $G_{\mathcal{M}}$ and $G_{\mathcal{R}}$ are related by the thermalization condition in \eq{thermal}, if follows that their Fourier transforms with respect to $(\tau-\tau')$ satisfies Eq.(\ref{ftrelation}). Hence, the Fourier transform of $G_{\mathcal{R}}$ is expected to satisfy:
\begin{align}\label{grft}
\tilde{G}_{\mathcal{R}}(\nu)=\tilde{G}_{\mathcal{M}}(\nu)-e^{-\pi|\nu|}\tilde{G}_{\mathcal{M}}^{(r)}(\nu)
\end{align}
Now, let us use Eq.(\ref{ftgm}) and Eq.(\ref{ftgmr}) in this equation to get
\begin{align}\label{grminusgm}
\tilde{G}_{\mathcal{R}}(\nu)=\frac{1}{\pi}\left[K_{i\nu}(e^{i\pi}\mu\rho_{<})-e^{-\pi|\nu|}K_{i\nu}(\mu\rho_{<})\right]K_{i\nu}(\mu\rho_{>})
\end{align} 
From the well known identity,
\begin{align}
K_{i\nu}(e^{i\pi}x)=e^{-\pi|\nu|}K_{i\nu}(x)-i\pi I_{-i|\nu|}(x);\qquad x>0
\end{align}
we can simplify Eq.(\ref{grminusgm}) to finally obtain:
\begin{align}
\tilde{G}_{\mathcal{R}}(\nu)=-iI_{-i|\nu|}(\mu\rho_{<})K_{i\nu}(\mu\rho_{>})
\end{align}
This matches exactly with the expression for $\tilde{G}_{\mathcal{R}}$ in Eq.(\ref{grft}) that we derived using a more direct method. 

\section{Discussion}

The physics in the Rindler frame, vis-a-vis the  Minkowski frame, has been investigated extensively in the literature. In view of this, it is useful to highlight new results and insights that this work offers. 
\begin{itemize}
	\item It is well known that Minkowski vacuum appears as a thermal state to a Rindler observer. A corollary of this result is that given the Feynman propagator $G_{\mathcal{R}}$, one can find $G_{\mathcal{M}}$ as a periodic sum of $G_{\mathcal{R}}$ (in the time variable $\tau$) with the period being $2\pi i$ (in appropriate units). It is natural to ask whether there exists an inversion of this procedure. That is,  given the expression for $G_{\mathcal{M}}$, can we find $G_{\mathcal{R}}$? In fact, one could ask whether such a result, if it exists, holds for a wider class of functions $\left\{F_{\mathcal{M}},F_{\mathcal{R}}\right\}$, wherein the former is a periodic sum of the latter with a period $2\pi i$? 
	We have explicitly shown that there exists an inverse transformation that retrieves $F_{\mathcal{R}}$ from its periodic sum $F_{\mathcal{M}}$, for a general class of functions $F_{\mathcal{R}}$. The Schwinger kernels $\left\{K_{\mathcal{M}},K_{\mathcal{R}}\right\}$ also form such a pair. Using the inverse procedure, we derived an integral representation for $K_{\mathcal{R}}$ in 1+1 dimensional flat spacetime, which is consistent with known results in the literature. (However, as far as we know, the explicit expression for $K_{\mathcal{R}}$ does not seem to exist in the previous literature.)
	
	\item In reference \cite{cr}, the authors derived an expression for $G_{\mathcal{R}}$ in terms if a curious integral transformation of $G_{\mathcal{M}}$. However, this result has not been explored further in the literature, in spite of the fact that $G_{\mathcal{R}}$ itself is a well-studied object. We have shown that this result is a direct consequence of the fact that $G_{\mathcal{M}}$ is a periodic sum of $G_{\mathcal{R}}$ with the period $2\pi i$. Specifically, it follows directly from the inverse transformation mentioned in the previous item. Hence, the result also holds true for the wider class of functions $\left\{F_{\mathcal{M}},F_{\mathcal{R}}\right\}$. In particular, the Schwinger kernel $K_{\mathcal{R}}$ can also be expressed as the integral transform of $K_{\mathcal{M}}$ in exactly the same manner.

	\item The Schwinger kernel for a scalar field, in the Euclidean sector, is just the heat kernel. For the two inequivalent vacua $\ket{\mathcal{M}}$ and $\ket{\mathcal{R}}$, the corresponding heat kernels are clearly different. However, it is well-known that: (a) the right(left) Rindler wedge, under Euclideanization with respect to $\tau$, maps to the whole of Euclidean flat space and (b) the heat kernel, in the Euclidean plane,  maps to $K_{\mathcal{M}}$ in the Lorentzian sector. The non-trivial task is to find the heat kernel, in the same Euclidean plane, that maps to $K_{\mathcal{R}}$. We explicitly demonstrate how this can be accomplished by solving the differential equation satisfied by the heat kernel, using appropriate mode functions. We show that, the existence of two inequivalent heat kernels, is related to the fact that Dirac delta function,  can be represented using two different sets of orthonormal mode functions; we obtain (i) (Euclidean) $K_{\mathcal{M}}$, when the modes used to represent the Dirac delta function are invariant under rotation by $2\pi$ and (ii) (Euclidean) $K_{\mathcal{R}}$, when the mode functions are not  invariant under rotation by any common finite angle.  
	
	\item The analytic continuation of the Euclidean polar coordinates --- which involves replacing $t_E\to -it$ and $\tau_E\to -i\tau$ in $x=\rho\cos\tau_E, t_E=\rho\sin\tau_E$ --- will  lead us only to the events in the right Rindler wedge. The question arises as to how one can extract the information contained in the other four wedges of the Lorentzian sector from the expression valid in the Euclidean sector. We have provided the four different analytic continuations of the Euclidean polar coordinates such that we can reach all the four wedges in the Lorentzian sector. The procedure is based on a simple unifying principle, viz that the analytic continuation should map Euclidean squared distance $\sigma_E^2$ to ($\sigma_M^2+i\epsilon$), with a positive, infinitesimal, imaginary part in the Lorentzian sector. We explicitly demonstrate that this procedure leads to the correct expressions for the propagators in the Lorentzian sector, even when the two events are in two different wedges. Again, as far as we know, this problem has not been explicitly addressed in the previous literature.
	
	This approach also allows us to discover relatively simple expressions for the temporal Fourier transform of the Minkowski and Rindler propagators. These two are connected by the relation in \eq{grft}, which we hope to study in detail in a future publication.

\end{itemize}      
Our procedure can be generalized to any bifurcate Killing horizons in \textit{curved} spacetime like e.g., the de Sitter horizon or black hole horizon. This implies that one can expect results similar to \eq{one} in much more general situations, even when the specific forms of $G_{\mathcal{M}}$ and $G_{\mathcal{R}}$ are not available. We hope to study this in detail in a future publication.

\section*{Acknowledgement}
KR is supported by Senior Research Fellowship (SRF) of the Council of Scientific \& Industrial Research (CSIR), India. Research of TP is partially supported by the J.C.Bose Fellowship of the Department of Science and Technology, Government of India.

\section*{Data Availability Statement}
Data sharing is not applicable to this article as no new data were created or analyzed in this
study.

\appendix
\numberwithin{equation}{section}
\section*{Appendix}

In the following sections we shall be mainly concerned with the derivations of certain results in the main body of this paper.  
\section{Derivation of Eq.(\ref{trigsum})}\label{appendixA}
In this section, we derive the result given in \eq{trigsum}. Let us start by denoting the desired sum by $S$. 
\begin{align}
S=\sum_{n=-\infty}^{\infty}e^{-|\tau_{E}+2\pi n|}
\end{align}
It is convenient to assume that: 
\begin{align}
-2\pi N<\tau_{E}<-2\pi (N-1)
\end{align}
where, $N$ is an integer (which can either be positive, negative or zero). Then the sum $S$ can be split into two parts as follows:
\begin{align}
S=\sum_{n=-\infty}^{N-1}e^{+\omega(\tau_{E}+2\pi n)}+\sum_{n=N}^{\infty}e^{-\omega(\tau_{E}+2\pi n)}
\end{align}
Each of the two summations on the right hand side of the above equation is a geometric series. Hence, using standard results, we get
\begin{align}
S&=\frac{e^{\omega[\tau_{E}+2\pi(N-1)]}}{1-e^{-2\pi\omega}}+\frac{e^{-\omega(\tau_E+2\pi N)}}{1-e^{-2\pi\omega}}
\end{align}
This can be further simplified into,
\begin{align}
S=\frac{\cosh[\omega\left\{\tau_{E}+\pi(2N-1)\right\}]}{\sinh(\pi\omega)}=\frac{\cosh[\omega\left(\underbar{$\tau$}_{E}-\pi\right)]}{\sinh(\pi\omega)}
\end{align}
where, $\underbar{$\tau$}_{E}=\tau_{E}\,\textrm{mod}\, 2\pi$.
Now, when $0<\tau_{E}<2\pi$, the above equation simplifies to
\begin{align}
S=\frac{\cosh[\omega(\tau_{E}-\pi)]}{\sinh(\pi\omega)}.
\end{align}
Similarity, when $-2\pi<\tau_{E}<0$, we get
\begin{align}
S=\frac{\cosh[\omega(\tau_{E}+\pi)]}{\sinh(\pi\omega)}.
\end{align}
Combining the last two equations, the expression for $S$, when $|\tau_{E}|<2\pi$ is given by
\begin{align}
S=\frac{\cosh[\omega(|\tau_{E}|-\pi)]}{\sinh(\pi\omega)}.
\end{align}
\section{Inverting the periodic summation}\label{AppendixB} 

In order to show that the transformations defined by \eq{k65} and \eq{k58} can be interpreted as inverse to each other, we have to show that one can retain an appropriate test function by implementing these transformations consecutively.

We will begin with the proof that \eq{k65} implies \eq{k58}.
Note that by definition, $F_{\mathcal{M}}(z+2\pi n i)=F_{\mathcal{M}}(z)$. Hence, for convenience, we assume that $-\pi<\textrm{Im}[z]<\pi$. Now, consider the RHS of \eq{k58}:
\begin{align}
\int_{\mathcal{C}'}\frac{du}{(i\pi)}\mathcal{H}(z;u)F_{\mathcal{R}}(u)&=\int_{\mathcal{C}'}\frac{du}{(i\pi)}\mathcal{H}(z;u)\left[\int_{\mathcal{C}}\frac{dv}{(i\pi)}\mathcal{G}(u;v)F_{\mathcal{M}}(v)\right]\\
&=\int_{\mathcal{C}}\frac{dv}{(i\pi)}\left[\int_{\mathcal{C}'}\frac{du}{(i\pi)}\mathcal{H}(z;u)\mathcal{G}(u;v)\right]F_{\mathcal{M}}(v)\label{1_2_step}
\end{align}
%%%
Consider the terms in the square bracket; using residue theorem, we get:
\begin{align}
\int_{\mathcal{C}'}\frac{du}{(i\pi)}\mathcal{H}(z;u)\mathcal{G}(u;v)&=\sum_{n=-\infty}^{\infty}\left(\frac{v}{v^2-(z+2\pi i n)^2}\right)\\
&=\frac{\sinh v}{2(\cosh v-\cosh z)}
\end{align}
Using this, \eq{1_2_step} can be simplified to
\begin{align}
\int_{\mathcal{C}}\frac{dv}{(i\pi)}\left[\int_{\mathcal{C}'}\frac{du}{(i\pi)}\mathcal{H}(z;u)\mathcal{G}(u;v)\right]F_{\mathcal{M}}(v)=\int_{\mathcal{C}}\frac{dv}{(2i\pi)}\frac{\sinh v}{(\cosh v-\cosh z)}F_{\mathcal{M}}(v)
\end{align}
It is convenient to deform $\mathcal{C}$ into the $\tilde{\mathcal{C}}$ given in Figure 6. There are three contributions to this integral: (1) the residue at $z$, (2) integral over the vertical line from $-i\pi$ to $i\pi$ and (3) integrals over the horizontal lines.
\begin{figure}
	\centering
	\includegraphics[scale=.3]{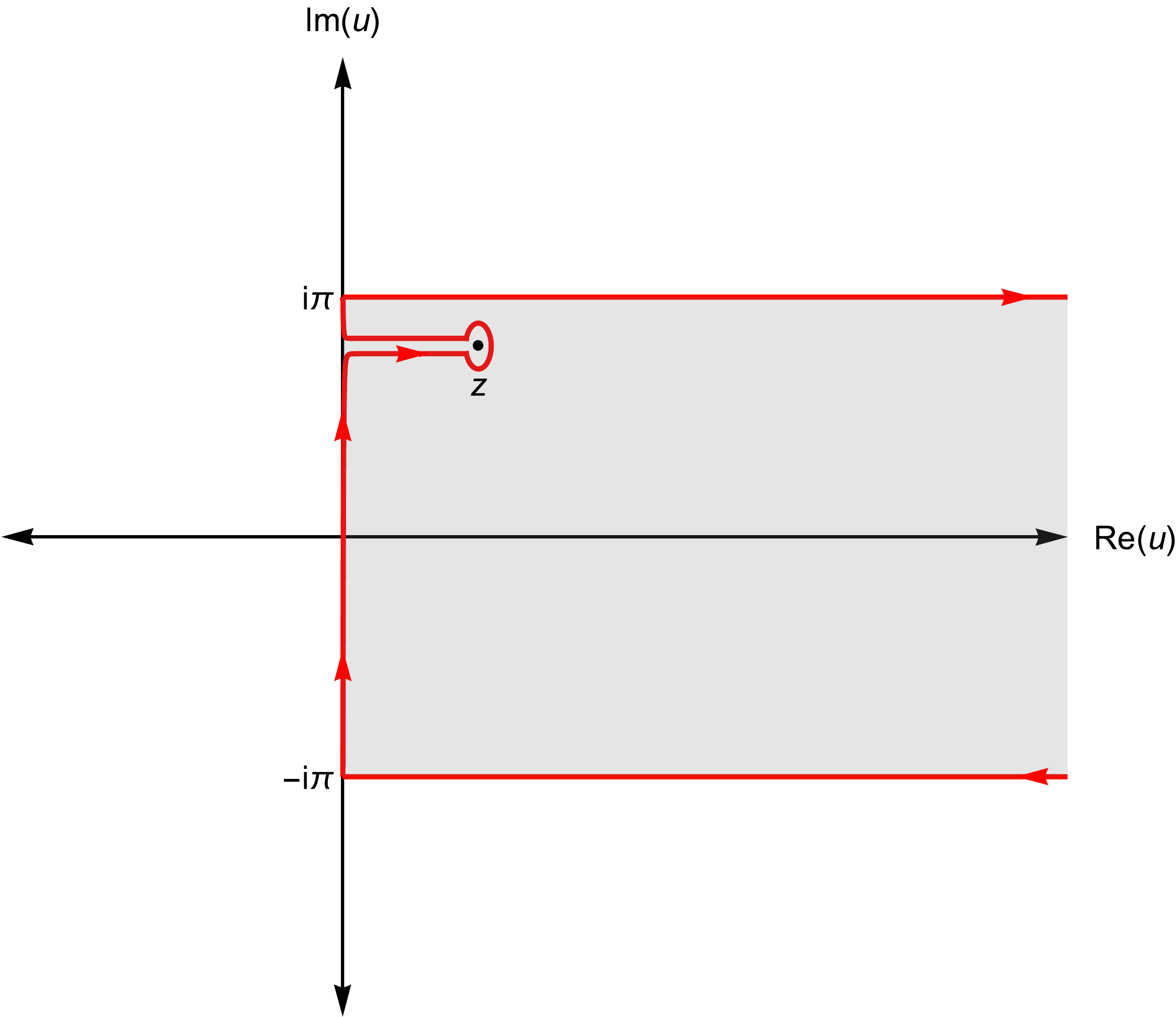}
	\caption{Contour $\tilde{\mathcal{C}}$.}
	\label{fig2}
\end{figure}
\begin{align}
\int_{\mathcal{C}'}\frac{dv}{(2i\pi)}\frac{\sinh v}{(\cosh v-\cosh z)}F_{\mathcal{M}}(v)&=F_{\mathcal{M}}(z)+\int_{-\pi}^{\pi}\frac{dy}{(2\pi)}\frac{i\sin y}{(\cos y-\cosh z)}F_{\mathcal{M}}(iy)\\\nonumber
&+\int_{\infty}^{0}\frac{dx}{(2i\pi)}\frac{\sinh x}{(\cosh x+\cosh z)}F_{\mathcal{M}}(x-i\pi)\\\nonumber
&+\int_{0}^{\infty}\frac{dx}{(2i\pi)}\frac{\sinh x}{(\cosh x+\cosh z)}F_{\mathcal{M}}(x+i\pi)
\end{align}
The second term in the first line vanishes, because $F_{\mathcal{M}}(iy)$ is even. From the pseudo periodicity and evenness of $F_{\mathcal{M}}$, we also have $F_{\mathcal{M}}(x-i\pi)=F_{\mathcal{M}}(x+i\pi)$. Hence, the sum of third and forth terms vanishes. Finally, we have
\begin{align}
\int_{\mathcal{C}'}\frac{du}{(i\pi)}\mathcal{H}(z;u)F_{\mathcal{R}}(u)&=F_{\mathcal{M}}(z).
\end{align}

Let us next show that \eq{k58} implies \eq{k65}. 
We start with the RHS of \eq{k65}.
\begin{align}
\int_{\mathcal{C}}\frac{du}{(i\pi)}\mathcal{G}(z;u)F_{\mathcal{M}}(u)&=\int_{\mathcal{C}}\frac{du}{(i\pi)}\mathcal{G}(z;u)\left[\int_{\mathcal{C}'}\frac{dv}{(i\pi)}\mathcal{H}(u;v)F_{\mathcal{R}}(v)\right]\\
&= \int_{\mathcal{C}'}\frac{dv}{(i\pi)}\left[\int_{\mathcal{C}}\frac{du}{(i\pi)}\mathcal{G}(z;u)\mathcal{H}(u;v)\right]F_{\mathcal{R}}(v)
\end{align}
Again, we will deform the contour $\mathcal{C}$ to $\tilde{\mathcal{C}}$, so that the term in the square bracket becomes,
\begin{align}
\int_{\mathcal{C}}\frac{du}{(i\pi)}\mathcal{G}(z;u)\mathcal{H}(u;v)&=\textrm{Residue term + vertical integral + horizontal integrals}
\end{align}
\textit{Note that the residue term is present only when $z$ lies in the shaded region of Figure 3}. In this case, we have
\begin{align}
\int_{\mathcal{C}}\frac{du}{(i\pi)}\mathcal{G}(z;u)\mathcal{H}(u;v)
&=\frac{\sinh v}{4(\cosh v-\cosh z)}+\\\nonumber
&+\int_{-\pi}^{\pi}\frac{dy}{\pi}\left[\frac{-iy}{(y^2+z^2)}\right]\left[\frac{\sinh v}{4(\cosh v-\cos y)}\right]\\\nonumber
&+\int_{\infty}^{0}\frac{dx}{(i\pi)}\left[\frac{(x-i\pi)}{(x-i\pi)^2-z^2}\right]\left[\frac{\sinh v}{4(\cosh v+\cosh x)}\right]\\\nonumber
&+\int_{0}^{\infty}\frac{dx}{(i\pi)}\left[\frac{(x+i\pi)}{(x+i\pi)^2-z^2}\right]\left[\frac{\sinh v}{4(\cosh v+\cosh x)}\right]
\end{align}
The second term in the last equation vanishes, since the integrand is odd. The third and forth term can be combined to get:
\begin{align}
&-\int_{0}^{\infty}\frac{dx}{(i\pi)}\left\{\frac{2i\pi(\pi^2+x^2+z^2)}{[\pi^2+(z-x)^2][\pi^2+(z+x)^2]}\left[\frac{\sinh v}{4(\cosh v+\cosh x)}\right]\right\}
&=-\int_{-\infty}^{\infty}  \frac{dx}{\pi^2+(z-x)^2}\left[\frac{\sinh v}{4(\cosh v+\cosh x)}\right]
\end{align}
Therefore, we have
\begin{align}
\int_{\mathcal{C}}\frac{du}{(i\pi)}\mathcal{G}(z;u)\mathcal{H}(u;v)&=\frac{\sinh v}{4(\cosh v-\cosh z)}-\int_{-\infty}^{\infty}  \frac{dx}{\pi^2+(z-x)^2}\left[\frac{\sinh v}{4(\cosh v+\cosh x)}\right]
\end{align}
So that,
\begin{align}
&\int_{\mathcal{C}'}\frac{dv}{(i\pi)}\left[\int_{\mathcal{C}}\frac{du}{(i\pi)}\mathcal{G}(z;u)\mathcal{H}(u;v)\right]F_R(v)\\
&=\int_{\mathcal{C}'}\frac{dv}{(i\pi)}\frac{\sinh v}{4(\cosh v-\sinh z)}F_R(v)-\int_{\mathcal{C}'}\frac{dv}{(i\pi)}\int_{-\infty}^{\infty}  \frac{dx}{\pi^2+(z-x)^2}\left[\frac{\sinh v}{4(\cosh v+\cosh x)}\right]F_R(v)\\
&=F_{M}(z)-\int_{-\infty}^{\infty}  \frac{dx}{\pi^2+(z-x)^2}F_{M}(x+i\pi)\\
&=F_{R}(z)
\end{align}

\section{The two Kernels: Technical details}\label{AppendixC}

\subsection{Derivation of $K^{Eu}_{\mathcal{R}}$}
We start with the expression for $K^{Eu}_{\mathcal{R}}$ given in \eq{rind_kernel}.
\begin{align}
K^{Eu}_{\mathcal{R}}&=\frac{1}{(2\pi s)}\exp\left(-\frac{r^2+r'^2}{4s}\right)\int_{0}^{\infty}d\omega\, I_{|\omega|}\left(\frac{rr'}{2s}\right)\cos\omega(\theta-\theta')
\end{align}
Recall that the modified Bessel function has the following contour integral representation.
\begin{align}\label{identity5v2}
I_{\nu}(z)=\frac{1}{2\pi i}\int_{\mathcal{C}}du\,e^{z\cosh u-\nu u};\qquad|\arg[z]|<\frac{\pi}{2}
\end{align}
where, the contour $\mathcal{C}$ is as given in Figure 3. Using this, the kernel $K_{\mathcal{R}}$ can be rewritten as
\begin{align}
K^{Eu}_{\mathcal{R}}&=\frac{1}{(2\pi s)}\exp\left(-\frac{r^2+r'^2}{4s}\right)\int_{0}^{\infty}d\omega \left[\frac{1}{2\pi i}\int_{\mathcal{C}}du\,e^{z\cosh u-\omega u}\right]\cos\omega\phi
\end{align}
where, $z=(rr')/(2s)$ and $\phi=|\theta-\theta'|$. Using the following identity,
\begin{align}
\int_{0}^{\infty}\cos\omega\phi e^{-\omega u}=\frac{u}{u^2+\phi^2};\qquad \left| \Im(\phi )\right| <\Re(u)
\end{align}
we can further simplify $K^{Eu}_{\mathcal{R}}$ to be,
\begin{align}
K^{Eu}_{\mathcal{R}}&=\frac{1}{(2\pi s)}\exp\left(-\frac{r^2+r'^2}{4s}\right)\left[\frac{1}{2\pi i}\int_{\mathcal{C}}du\,e^{z\cosh u}\left(\frac{u}{u^2+\phi^2}\right)\right]
\end{align}
Let us now assume that $\phi>\pi$, and simplify this expression, which can later be analytically continued for $\phi<\pi$. The following identity may be used,
\begin{align}
\int_{0}^{\infty}d\nu\sin(\nu u) e^{-\nu \phi}=\frac{u^2}{u^2+\phi^2}; \qquad \left| \Im(u)\right| <\Re(\phi )
\end{align}
When $\phi>\pi$, the condition for the above integral is satisfied, so that $K^{Eu}_{\mathcal{R}}$ may be written as
\begin{align}
K^{Eu}_{\mathcal{R}}&=\frac{1}{(2\pi s)}\exp\left(-\frac{r^2+r'^2}{4s}\right) \left[\frac{1}{2\pi i}\int_{\mathcal{C}}du\,e^{z\cosh u}\left(\int_{0}^{\infty}d\nu\sin(\nu u) e^{-\nu \phi}\right)\right]
\end{align}
Again, using the integral representation of $I_{\nu}$, we obtain
\begin{align}
K^{Eu}_{\mathcal{R}}=\frac{1}{(2\pi s)}\exp\left(-\frac{r^2+r'^2}{4s}\right) \int_{0}^{\infty}d\nu\, e^{-\nu \phi}\left(\frac{I_{-i\nu}(z)-I_{i\nu}(z)}{2i}\right)
\end{align}
Recall that the MacDonald $K_{\mu} (z)$ is given by
\begin{align}
K_{\mu}(z)= \frac{\pi}{2}\left(\frac{I_{-\mu}(z)-I_{\mu}(z)}{\sin(\pi\mu)}\right)
\end{align}
Hence,
\begin{align}
K^{Eu}_{\mathcal{R}}&=\frac{1}{(2\pi s)}\exp\left(-\frac{r^2+r'^2}{4s}\right) \int_{0}^{\infty}\frac{d\nu}{\pi} e^{-\nu \phi}\sinh(\pi\nu)K_{i\nu}\left(\frac{rr'}{2s}\right)
\end{align}

\subsection{$K^{Eu}_{\mathcal{M}}$ in terms of MacDonald function} 

Let us start with the following representation of $K^{Eu}_{\mathcal{M}}$ that follows from \eq{k17}.
\begin{align}
K^{Eu}_{\mathcal{M}}=\frac{1}{(4\pi s)}\exp\left(-\frac{r^2+r'^2}{4s}\right)\sum_{m=-\infty}^{\infty} I_{m}\left(z\right)e^{im\phi}
\end{align}
where, $z=(rr'/2s)$ and $\phi=|\theta-\theta'|$. Using the contour representation of $I_{\nu}$, we have
\begin{align}\label{kmint}
K^{Eu}_{\mathcal{M}}&=\frac{1}{(4\pi s )}\exp\left(-\frac{r^2+r'^2}{4 s }\right)\sum_{m=-\infty}^{\infty}\left[\frac{1}{2\pi i}\int_{\mathcal{C}}du\,e^{z\cosh u-m u}\right]e^{im\phi}
\end{align}
The following identity is useful,
\begin{align}
-1+\sum_{m=0}^{\infty}2\cos(m\phi) e^{-mu}&=\frac{\sinh (u)}{\cosh (u)-\cos (\phi )};\qquad \Re[u]\geq 0\\
&=\sum_{n=-\infty}^{\infty}\frac{2u}{u^2+(\phi+2\pi n)}\\
&=2\sum_{n=-\infty}^{\infty}\left[\int_{0}^{\infty}\sin(\nu u)e^{-\nu|\phi+2\pi n|}\right]\\
&=2\int_{0}^{\infty}d\nu\sin(\nu u)\left[\frac{\cosh\left\{\nu(\pi-\phi)\right\}}{\sinh(\pi\nu)}\right];\qquad 0<\phi<\pi.
\end{align}
Using this in \eq{kmint} we get,
\begin{align}
K^{Eu}_{\mathcal{M}}&=\frac{1}{(2\pi s )}\exp\left(-\frac{r^2+r'^2}{4 s }\right)\int_{0}^{\infty}d\nu\frac{\cosh\left[\nu(\pi-\phi)\right]}{\sinh(\pi\nu)}\left[\frac{1}{2\pi i}\int_{\mathcal{C}}du\,e^{z\cosh u}\sin(\nu u)\right]
\end{align}
Again using the integral representation of $I_{\nu}$ and the relation between $K_{\nu}$ and $I_{\nu}$, we get
\begin{align}
K^{Eu}_{\mathcal{M}}&=\frac{1}{(2\pi s )}\exp\left(-\frac{r^2+r'^2}{4 s }\right)\int_{0}^{\infty}\frac{d\nu}{\pi}\cosh[\nu(\pi-\phi)]K_{i\nu}\left(\frac{rr'}{2s}\right)
\end{align}

\section{Details of analytic continuation} \label{appendixd}
In this section we will discuss some technical details of our recipe for analytic continuation in Table \ref{tab:table1}. Here, we explicitly show that under these transformations (i) $\sigma_{E}^2\rightarrow\sigma^2+i\epsilon$ and (ii) the analytic continuation of $\Theta_{E}$ that follows from Eq.(\ref{defTheta}) matches exactly with that in given Table \ref{tab:table1}.   

\subsection{RR: both events on the R wedge}
The transformations, in this case, are given by
\begin{align}
(r,\theta)\rightarrow(\rho,i\tau e^{-i \epsilon});&&(r'\theta')\rightarrow(\rho',i\tau'e^{-i\epsilon}).
\end{align}
Therefore, to leading order in $\epsilon$, the invariant distance squared is given by
\begin{align}
\sigma_{E}^2\rightarrow \left[\rho ^2-2 \rho  \rho ' \cosh \left(\tau -\tau '\right)+\left(\rho '\right)^2\right]+2 i \epsilon \rho    \rho ' \left(\tau -\tau '\right) \sinh \left(\tau -\tau
'\right)+O\left(\epsilon ^2\right).
\end{align}
Clearly the imaginary part of the right hand side is positive. Now, to find the analytic continuation of $\Theta_{E}$, let us first look at the expression for $Z$.
\begin{align}
Z&=\cosh(e^{-i\epsilon|\tau-\tau'|})
\end{align}
This implies that
\begin{align}
1-Z^2&=-\sinh ^2\left(\tau -\tau '\right)+i \epsilon  \left(\tau -\tau '\right) \sinh \left[2 \left(\tau -\tau '\right)\right]+O\left(\epsilon ^2\right)
\end{align}
Since $1-Z^2$ has a small positive imaginary component, the square root in the expression for $\Theta_{E}$ will evaluate to $\sqrt{1-Z^2}=i|\sinh\left(\tau -\tau '\right)|$. Hence, the final expression for $\Theta_{E}$ simplifies to:
\begin{align}
\Theta_{E}\rightarrow\frac{\pi}{2}+i\log(ie^{|\tau-\tau'|(1-i\epsilon)})=i|\tau-\tau'|+0^{+}
\end{align}
\subsection{RF: one event on R and the other on F wedge}
According to our recipe, the transformation of points in this case is given by
\begin{align}
(r,\theta)\rightarrow(\rho_{R},i\tau_{R});&&(r',\theta')\rightarrow(i\rho_{F},i\tau_{F}+\frac{\pi}{2}+\epsilon).
\end{align}
Hence, $\sigma_{E}^2$ transforms to
\begin{align}
\sigma_{E}^2\rightarrow\left[-\rho _F^2-2 \rho _F \rho _R \sinh \left(\tau _F-\tau _R\right)+\rho _R^2\right]+2 i \epsilon  \rho _F \rho _R \cosh \left(\tau _F-\tau _R\right)+O\left(\epsilon ^2\right).
\end{align}
Once again, we see that the imaginary part of the right hand side is positive. The biscalar $Z$ in this case turns out to be: 
\begin{align}
Z&=-i\sinh\left[(\tau_{F}-\tau_{R})-i\epsilon\right]
\end{align}
It then follows that
\begin{align}
1-Z^2&=\cosh\left[(\tau_F-\tau_{R})-i\epsilon\right]
\end{align} 
Therefore, the analytic continuation of $\Theta_{E}$ becomes;
\begin{align}
\Theta_{E}\rightarrow\frac{\pi}{2}+i\log(e^{(\tau_F-\tau_R)-i\epsilon})=i(\tau_F-\tau_{R})+\frac{\pi}{2}+\epsilon.
\end{align}
Notice that there is no `modulus' on $\tau_{F}-\tau_{R}$.
\subsection{FF: both events on the F wedge}
Following our prescription in Table \ref{tab:table1}, the analytic continuation of coordinates is given by:
\begin{align}
(r_{<},\theta)\rightarrow\left(-e^{i\epsilon}i\rho_{<},i\tau+\frac{\pi}{2}\right);&&(r_{>},\theta')\rightarrow \left(i\rho_{>},i\tau'-\frac{\pi}{2}\right).
\end{align}
Under this transformation, the square of distance $\sigma_{E}^2$ continues to:
\begin{align}
\sigma_{E}^2\rightarrow\left(2 \rho _{>} \rho _l \cosh \left(\tau -\tau '\right)-\rho _{>}^2-\rho _{<}^2\right)-2 i \epsilon  \rho _{<} \left(\rho _{>} \cosh \left(\tau -\tau '\right)-\rho
_{<}\right)+O\left(\epsilon ^2\right)
\end{align}
Since, by definition $\rho_{>}\geq\rho_{<}$, we find that the imaginary part of the right hand side is positive. The biscalar $Z$ in this case becomes:
\begin{align}
Z&=-\cosh(\tau-\tau'),
\end{align}
which implies
\begin{align}
1-Z^2&=-\sinh^2(\tau-\tau').
\end{align}
Therefore, $\sqrt{1-Z^2+i0^{+}}=i\sinh|\tau-\tau|$. Finally, the expression for $\Theta_{E}$ reduces to;
\begin{align}
\Theta_{E}\rightarrow\frac{\pi}{2}+i\log(-ie^{-|\tau-\tau'|})=-i|\tau-\tau'|+\pi
\end{align}

\bibliography{rindlervac}

\begin{thebibliography}{10}
\bibitem{note2}
 If we consider a massive scalar field then the corresponding kernel will
  involve the mass $m$ and is related to massless heat kernel by the simple
  relation: $K_{m\ne 0} (x,x_0; s) = e^{-im^2 s} \ K_{m=0} (x, x_0; s)$ so that
  it is enough to concentrate on the massless case, most of the time.
\bibitem{note3}
 Defining the Schwinger kernel through an explicit solution of Eq. (\ref{solk})
  is somewhat different --- conceptually and technically --- from defining the
  kernel simply as a solution to the partial differential equation Eq.
  (\ref{pdektp}). In the latter approach, one has a certain level of ambiguity
  in the solution since one could add any solution to the homogeneous part of
  the equation. This ambiguity is removed when we define the kernel directly
  through Eq. (\ref{solk}).) Therefore one cannot claim that the difference
  between Rindler vacuum and the Minkowski vacuum can be taken care of by
  adding a suitable solution to the homogeneous part of the differential
  equation in Eq. (\ref{pdektp}).
\bibitem{note4}
 In fact, the result in \eq{one} holds for a wider class of functions, viz.,
  all even functions $F_{\mathcal{R}}(\tau)$. Such a function can be expressed
  as a superposition of $\exp(-i\omega|\tau|)$ with positive and negative
  $\omega$ with an weightage factor that is even in $\omega$. We will comment
  on this generalization towards the end.
\bibitem{note6}
 If one approaches the problem via standard canonical quantization, this
  question has a seemingly natural answer. We first obtain the relevant
  positive frequency modes in Rindler and Minkowski coordinates and define the
  corresponding Rindler and Minkowski vacua. The propagators are then defined
  as expectation values of products of field operators in these two vacua.
  However, we are approaching it from the point of view of solutions to a
  particular differential equation. It is then necessary to understand how the
  distinction between the two vacua is encoded in these solutions.
\bibitem{note7}
 In the context of QM in a 2-dimensional plane, one would not dream of writing
  the Dirac delta function with just $\delta(\theta-\theta_0)$. In QM, the
  kernel evolves an initial wave function to a final wave function; in fact
  $K(x,x_0;s)$ is indeed the wave function for a particle at time $s$ if it
  started out as a delta function, $\delta(x,x_0)$ at time $s=0$. So to justify
  the choice of the delta function with a factor $\delta(\theta-\theta_0)$ one
  should be able to interpret it as the initial wave function of a particle in
  the plane. This is tricky because one would expect the wave functions to be
  single-valued. Since $\theta$ and $\theta+2\pi$ represent the same physical
  point, the function $\delta(\theta-\theta_0)$ cannot be an acceptable choice
  for the initial wave function. Therefore the kernel $K_{n=0}$ never arises in
  the context of, say, normal QM.
\bibitem{note8}
 Put $y=\cos\theta$, where $0\leq\theta\leq\pi$, so that
  $iy+\sqrt{1-y^2}=ie^{-i\theta}$. Therefore,
  $\cos^{-1}(y)=\frac{\pi}{2}-i\log(ie^{-i\theta})=\theta$.
\bibitem{note1}
 For large $|u|$, the leading order asympotic expansion for $I_{u}(z)$ is given
  by $I_{u}(z)\sim[\Gamma(1+u)]^{-1}\left(\frac{z}{2}\right)^{u}$. This implies
  that $$I_{i\nu}(\mu\rho_{<})K_{i\nu}(\mu\rho_{>})\sim \frac{1}{2\nu
  i}\left(\frac{\rho_{>}}{\rho_{<}}\right)^{-\textrm{Im}[\nu]+i\textrm{Re}[\nu]};\qquad
  \textrm{Im}[\nu]\rightarrow-\infty. $$ Clearly the R.H.S vanishes as
  $\textrm{Im}[\nu]\rightarrow-\infty$ when $\rho_{>}>\rho_{<}$.
\end{thebibliography}

\end{document}